\documentclass[prb,superscriptaddress,twocolumn,amsmath,nofootinbib,longbibliography]{revtex4-1}
\usepackage{amsmath,graphicx,bm,wasysym,color}
\usepackage{amsmath,amsfonts,amsthm,amssymb,amsbsy,bbold,mathrsfs}

\newcommand{\sect}{\section}
\newcommand{\subsect}{\subsection}
\begin{document}
\title{Anomalous localization at the boundary of an interacting topological insulator}
\author{Itamar Kimchi}
\email{ikimchi@gmail.com}
\affiliation{Department of Physics and Center for Theory of Quantum Matter, University of Colorado, Boulder, CO 80309, USA}
\affiliation{JILA, NIST and Department of Physics, University of Colorado, Boulder, CO 80309, USA}
\author{Yang-Zhi Chou}
\affiliation{Condensed Matter Theory Center and the Joint Quantum Institute, Department of Physics, University of Maryland, College Park, MD 20742, USA}
\affiliation{Department of Physics and Center for Theory of Quantum Matter, University of Colorado, Boulder, CO 80309, USA}
\author{Rahul M. Nandkishore}
\affiliation{Department of Physics and Center for Theory of Quantum Matter, University of Colorado, Boulder, CO 80309, USA}
\author{Leo Radzihovsky}
\affiliation{Department of Physics and Center for Theory of Quantum Matter, University of Colorado, Boulder, CO 80309, USA}


\begin{abstract}
The boundary of a topological insulator (TI) hosts an anomaly restricting its possible phases: e.g.\ 3D strong and weak TIs maintain surface conductivity at any disorder if symmetry is preserved on-average, at least when electron interactions on the surface are weak. 
However the interplay of strong interactions and disorder with the boundary anomaly has not yet been theoretically addressed. 
Here we study this combination for the edge of a 2D TI and the surface of a 3D weak TI, showing how it can lead to an ``Anomalous Many Body Localized'' (AMBL) phase that preserves the anomaly. 
We discuss how the anomalous Kramers parity switching with $\pi$ flux arises in the bosonized theory of the localized helical state. 
The anomaly can be probed in localized boundaries by electrostatically sensing nonlinear hopping transport with e/2 shot noise. 
Our AMBL construction in 3D weak TIs  fails for 3D strong TIs, suggesting that their anomaly restrictions are distinguished by strong interactions.
\end{abstract}
\maketitle

%
%

\sect{Introduction: topology, correlations and disorder}

\textit{Motivation.} 
Topological insulator phases of weakly interacting electrons\cite{Mele2005,Mele2005a,Mele2007,Balents2007,Roy2009,Kane2010,Zhang2011} are known to exhibit anomalies on their boundaries. The anomalies imply that any phase that can emerge on the boundary has a response that cannot be emulated by a local lattice model with standard symmetries. This remains true even when the symmetry that protects the topological phase is locally broken, spontaneously or with symmetry-breaking disorder, as long as the symmetry broken order parameter is zero on average.
\cite{Stern2012,FuKaneTopology2012,Moore2012,Akhmerov2014,Mudry2015} 
 For example (as we elaborate on below) the surface of a 3D strong or weak topological insulator is known to remain metallic and forbid localization (at least for weakly interacting electrons) if the symmetry broken domains occur with equal probability, through a mechanism that can be viewed as 2D percolation of domain walls. 
 However the combination of arbitrary disorder or local symmetry breaking with arbitrarily strong interactions has not been as well explored, and in particular their consequences for the protected anomaly have not yet been understood.

\begin{figure}
	\includegraphics[width=0.45\textwidth]{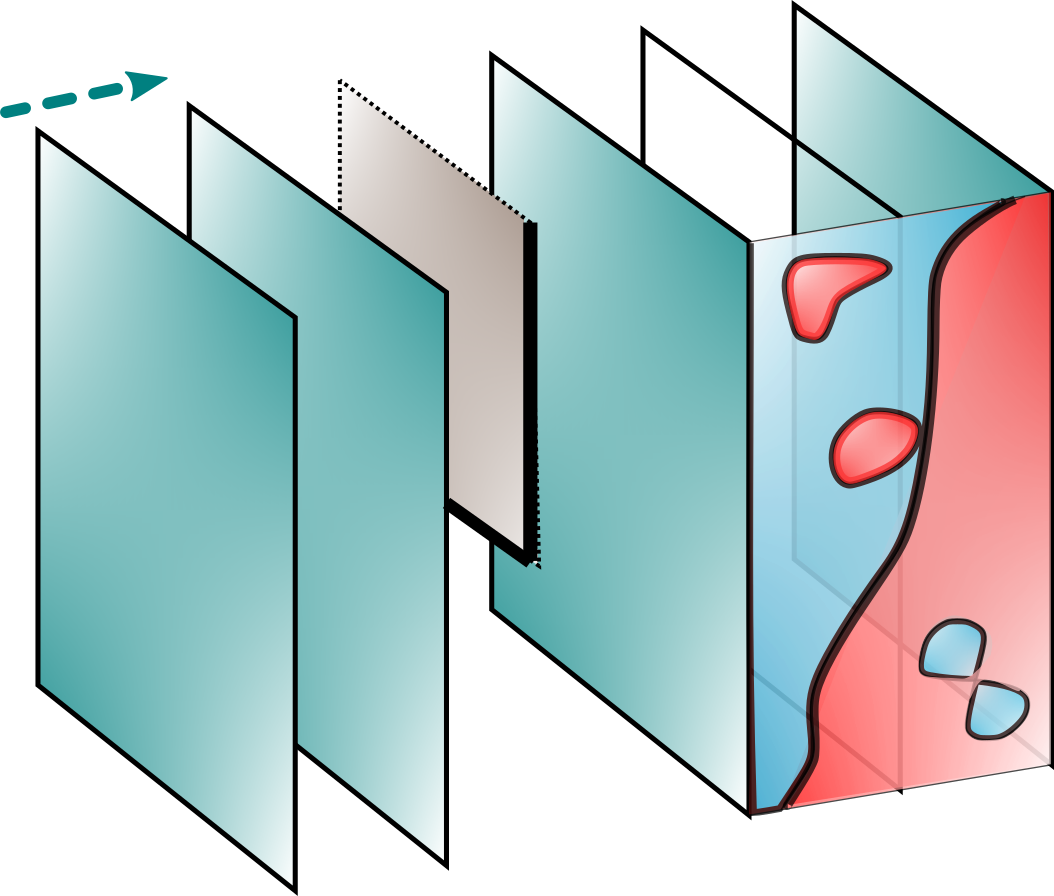}
	\caption{Three dimensional weak TI and its AMBL surface. The weak TI is pictured as a stack of 2D TIs with a translation vector (dashed arrow) defining its $Z_2$ symmetry. Helical Luttinger Liquids (thick black lines) are bound to dislocations \cite{Vishwanath2009} (gray partial-plane), and are also bound to domain walls between $Z_2$-translation-broken hybridization domains (blue and red, shown schematically upon coarse graining)  on side surfaces. 
	Weak $Z_2$-symmetry-breaking disorder that preserves $Z_2$ on average produces a helical network with a spanning helical mode and sparse junctions. Stronger symmetry-preserving disorder then allows this spanning helical LL to localize as in Fig.~\ref{fig:dw}. }
	\label{fig:wti}
\end{figure}

\textit{Present work.} 
Here we study the aforementioned constraints from a topological anomaly for the case of arbitrary disorder and interactions, focusing on electron 2D TIs and 3D weak TIs, and show that in these cases (but not for 3D strong TIs), combining interactions with disorder leads to an unusual localized phase that is not allowed with either interactions or disorder alone. We derive constraints for the interacting many-body-localized (MBL) phases that can occur and show how they differ from conventional MBL phases by virtue of exhibiting the anomaly, motivating the terminology ``Anomalous-MBL'' (AMBL) phases. 
In an l-bit description, the AMBL phase has at least one ``bit" that is a non-localizable degree of freedom (hence not a complete set of l-bits), which allows for the anomaly response.
Working from a recent result using bosonization to construct a localized spin-glass edge of a 2D TI \cite{Chou2018}, we show in detail how bosonization captures the anomaly in an anomalous-MBL fashion. Here the non-localizable bit is due to the spontaneous symmetry breaking of time reversal, and the localized objects carry charge e/2. 
We apply these ideas to construct a localized surface of a 3D \textit{weak} TI (Fig.~\ref{fig:wti}), and discuss the phenomenology of the localized state including conventional nonlinear IV and anomalous e/2 shot noise. As we will see construction does not extend to a 3D \textit{strong} TI, thus providing a previously unappreciated distinction between 3D strong and weak TIs that manifests only in the presence of both disorder and strong interactions. 


\textit{Theoretical context.} 
Our present setting should be distinguished from two types of phases that are known to arise in the phase diagram of topological insulator surfaces.  Interactions allow the surface of a 3D TI to enter a surface topological order phase with fractionalized excitations that transform anomalously under certain symmetries. \cite{Senthil2013,Vishwanath2014b,Qi2013} 
Interactions can also lead to a state with a net (nonzero on-average) symmetry breaking order parameter, e.g.\ a ferromagnet in the strong TI case; such a surface  appears trivial though the anomaly is still manifested in defects of the symmetry-breaking order such as domain walls.
Here we are concerned strictly with states that do not show surface topological order and also do not show net on-average symmetry breaking, but that can show local symmetry breaking with zero average order parameter.
 
To study the phases that can arise on a TI surface upon introducing interactions and disorder it is important to prevent the bulk from undergoing a phase transition. A theoretically precise setting to avoid a bulk phase transition is to consider interactions and disorder that appear only on the surface. 
The results will continue to hold as interactions and disorder are increased in the bulk as long as they remain below the bulk gap. Note also that experimentally it is quite reasonable to take disorder to be substantially stronger on the surface than in the bulk; it is also plausible for interactions to differ based on screening or superexchange via a substrate on a boundary compared to the bulk.

Early on, the topological protection for boundaries of topological insulators was studied for the case of arbitrary disorder  using various tools constructed for noninteracting electrons.  This led to the classification of topological insulators and superconductors \cite{Ludwig2008,Ludwig2010} and associated results applicable to this weakly interacting limit. 
Both strong and weak 3D TIs have  been shown to be protected against localization from symmetry-breaking disorder, in a weakly interacting limit, via parallel mechanisms. \cite{Stern2012,FuKaneTopology2012,Moore2012,Akhmerov2014,Mudry2015} (Recall that the protecting symmetry for strong TIs is time reversal symmetry; while for weak TIs, denoted by a triplet of $Z_2$ indices multiplying reciprocal lattice vectors whose sum corresponds to a stacking orientation of 2D TIs, it is the $Z_2$ subgroup of the translation symmetry along the stacking vector.)
For example, creating local symmetry-breaking (via disorder or interactions) to the surface of 3D strong or weak electronic TIs leads to domains and a network of domain walls that are conducting. As long as the symmetry is preserved on average (``statistical symmetry", i.e.\ the local symmetry-breaking order parameter can be viewed as drawn from a distribution that preserves the symmetry and is short ranged correlated, hence also the two domains appear with equal probability), this conducting network percolates.  \cite{Stern2012,FuKaneTopology2012,Moore2012,Akhmerov2014,Mudry2015} 

Though the constructions have all relied on theoretical machinery based on noninteracting electrons, at least in some cases it seems natural to expect the conclusions to persist to strong interactions. For example, in the Fu-Kane approach\cite{FuKaneTopology2012}, the protection of metallicity on both the 3D weak TI and 3D strong TI surface is given by a $Z_2$ index associated with the vortex fugacity in the strong/weak TI bulk. This vortex $Z_2$ index naively appears to be equivalent to the $Z_2$  index denoting the topological state which is known in both cases to be preserved by arbitrary interactions. But is the vortex $Z_2$ index and associated protected metallicity also necessarily preserved by arbitrary interactions?

The generic case of strong interactions and strong disorder has been explored for 2D TI edges through 1D  bosonization, \cite{Zhang2006,Moore2006,Chou2018}
leading (for appropriate disorder and interaction strengths) to a zero-temperature state which breaks time-reversal locally but preserves it on average and may naively appear to be a many-body localized (MBL)\cite{Huse2015,Serbyn2019} phase. The phase can be thought of as a spin glass, where domain walls (kinks and anti-kinks) form localized puddles of charge-e/2 fermions (solitons).  It is a non-Fermi glass \cite{Gopalakrishnan2017} and is gapless  similarly to the conventional Bose glass \cite{Fisher1989}. Taken at face value, within bosonization the phase and its entire spectrum (within the low energy theory) appears to be fully localized.  
Indeed the modern understanding of MBL with a locator expansion and the associated picture of l-bits\cite{Abanin2013,Oganesyan2014,Imbrie2016}, which are a complete set of mutually commuting local conserved  quantities, gives strong constraints. Resolving how to modify this conventional l-bit MBL picture in the presence of the anomalies  is the essence of the main result in Section \ref{sec:2dti} below, where we also make the notion of a regularized Hilbert space from bosonization more precise.

Here we consider TI systems in class AII with time reversal symmetry $T$ and electron charge $U(1)$ conservation. We shall refer to the 1D edge as an anomalous ``helical Luttinger liquid (LL)'' to distinguish it from non-anomalous pure-1D spin-orbit-coupled systems. Elsewhere 2D TIs have also been considered with an additional spin U(1) symmetry of $S^z$ conservation \cite{Yudson2013,Loss2017}.
Localized boundaries of topological phases in symmetry classes other than AII have been previously studied \cite{Yuzbashyan2012,Chou2014}, including recently \cite{Radzihovsky2019} via localization of a helical network  on the surface of class CII spin chiral topological insulators. Class AII shows distinct anomalies. Note that in class CII, a surface helical network arises when one of the CII symmetry generators is locally broken while the other remains an exact symmetry; without this additional symmetry constraint the CII TI surface  can be fully localized and fully gapped already in the free fermion case with symmetry-breaking-mass textures whose net order parameter vanishes, in contrast to the physics we consider below.

\sect{Anomaly and localization on 2D TI edges}
\label{sec:2dti}
\subsect{General argument for impossibility of full symmetry-preserving MBL in a 2D TI boundary}
To see the restrictions on driving the 1D helical LL into a symmetry-preserving MBL phase, even with arbitrarily strong electron correlations and disorder, consider the manifestation of the anomaly of 2D TIs on their 1D helical edge. 
As we elaborate on below, the essential Eq.~known from previous work \cite{Lee2008,Zhang2008} is the response to boundary condition twists in the relation between the system's many-body time-reversal Kramers parity $T^2$ and electron fermion parity,
\begin{equation}T^2 = (-1)^{n_f} \nu, \ \nu \equiv e^{i \Phi}
\label{eq:nu}
\end{equation}
in the anomalous 1D helical LL, where $\Phi=0,\pi$ is the (time-reversal-invariant) flux enclosed by the 1D helical edge, or equivalently $\nu = -1$ for anti-periodic boundary conditions and $+1$ for periodic.
A second anomalous feature of the 2D TI edge \cite{Zhang2008a} is the appearance of $\pm$e/2 electric charges at the pointlike domain wall between opposite time-reversal-broken domains; this feature appears straightforwardly in the localized states we discuss here, but we return to it in Section~\ref{sec:halfcharge}  when discussing observable phenomenology.
Below we first remind the reader of the physical content in Eq.~(\ref{eq:nu}), and then derive our various results for how Eq.~(\ref{eq:nu}) is realized and resolved in the anomalous localization case.

The reader new to  Eq.~(\ref{eq:nu}) can contrast this with a pure 1D system, where the many-body action of $T^2$ is just given by the fermion parity. With the anomaly here, in contrast, $T^2$  gains an additional sign change with a $\pi$ twist in the boundary conditions. 
In the bulk of the 2D TI, the anomaly can be probed \cite{Lee2008,Zhang2008} by threading a $\pi$ flux and observing that a Kramers doublet appears near the threaded flux.  Eq.~(\ref{eq:nu}) is the manifestation of this 2D bulk anomaly on the 1D edge: the $\pi$ flux, seen by the edge as a twist of the electron wavefunction boundary conditions to anti-periodic boundary conditions, must also turn the many body edge state into a Kramers doublet.    

This factor $\nu$ of the helical LL anomaly is well known \cite{Lee2008,Zhang2008} but for completeness let us recall simple arguments for its construction. 
Though here we are concerned with $Z_2$ 2D and 3D topological insulators, the factor $\nu$ can be seen most easily by adding  for the moment an additional U(1) spin conservation symmetry.  Then the 2D TI is two copies of IQHE with opposite Chern-Simons responses for opposite spins, and one can simply use Laughlin's argument of $2 \pi$ flux insertion through a solenoid (which builds a circulating EMF and transfers an electron from the edge). Tracking spin here shows that threading $2\pi$ flux transfers a spin-1 quantum number to the vicinity of the threaded flux, while threading $\pi$ flux binds a spin quantum number of ${\pm}1/2$. Without the additional U(1) conservation this transferred spin-1/2 is nothing more than a Kramers doublet. On the 1D edge, the U(1) spin conservation symmetry together with charge conservation implies that both species are conserved: this is exactly the setting for the 1+1D chiral anomaly. The chiral anomaly is manifested by coupling the fermions to a background gauge field. Inserting $\pi$ flux creates a spin current and pumps an imbalance $n_\uparrow - n_\downarrow = {\pm}1$, corresponding to a single fermion filling the up/down state, hence a Kramers doublet with spin-1/2. An alternative easy way to see the response of the edge, without a U(1) spin symmetry, is just by filling the noninteracting band structure of the edge helical LL: antiperiodic boundary conditions shift all momentum states and thereby require changing the fermion occupancy of the $k{=}0$ state which is a Kramers doublet. This particular manifestation, known as level pair switching in the non-interacting case, no longer directly applies to the strongly correlated setting, but the associated Kramers-doublet-switching of  Eq.~(\ref{eq:nu}) is robust to adding interactions.

Our key observation is that the anomaly features, including Eq.~(\ref{eq:nu}), persist in the presence of locally symmetry breaking disorder and interactions. 
Thus we now derive a restriction on localization that arises from the observation of  Eq.~(\ref{eq:nu}), namely that the Hilbert space of the 1D helical LL is necessarily sensitive to a change in its boundary conditions. The sensitivity is that changing boundary conditions from periodic to antiperiodic switches the Kramers doublet characteristic of the system. Intuitively it is reasonable that a response to a twist in the boundary conditions implies that the system cannot be strictly localized. To see this explicitly, consider the time reversal operation acting on the many body system $T$; for clarity let us now write it with a subscript as  $T_{MB}$. For certain classes of quantum wavefunctions that can be written as a product of local Hilbert spaces, in particular including l-bit MBL states, the action of $T_{MB}$ can be factorized into a product of local (but not necessarily on-site) time-reversal operations\cite{Vishwanath2015,Vasseur2016}. As we shall see below, here we need only rely on such a decomposition for the squared time reversal operator, $T^2_{MB}$. Let us express it as a product of operations on individual l-bits, in a putative MBL state:
\begin{equation}
T^2_{MB} = \prod_l T^2_l.
\end{equation}

Now consider the insertion of $\pi$ flux enclosed by the periodic 1D edge and its relation to $T_{MB}$. The state with $\pi$ flux still preserves time-reversal. The flux is represented as a twist in the periodic boundary conditions of the 1D system. As usual there are many ways to implement this twist that are all equivalent up to the gauge choice of locally redefining the electron operator U(1) phase, representing gauge choices of the gauge-invariant $\pi$ flux. Some gauge choices explicitly preserve time-reversal: these involve a sign difference $(-1)$ for the electron hopping across some chosen bond. (Viewing the 1D helical LL as the edge of a 2D TI, the sign offset bonds extend in a ray through the TI bulk.) Other gauge choices allow the $\pi$ flux to be implemented non-locally as a uniform $1/L$ perturbation  (where $L$ is the system size) in the hopping phase across each bond, at the cost of losing the original $T_{MB}$ symmetry, instead replacing it with a modified symmetry operation $T'_{MB}$ which is related to $T_{MB}$ by the gauge phase transformation. But the phase transformation is purely unitary so $T^2_{MB}$ remains a symmetry and is unmodified. Now we can simply observe that a small $1/L$  perturbation cannot modify the $T^2_l$ action on any single l-bit (at least if the l-bits have finite support -- see below). 
This result is easy to see only for some gauge choices but as a gauge-invariant statement it is necessarily a general result on the impossibility of changing  $T^2_{MB}$ with flux insertion. 
Strictly speaking this result assumes that at some parameter point within the putative MBL phase, the l-bits can be reached from the original lattice site by a unitary transformation whose spatial extent has only finite support. Generic l-bits  involve a unitary transformation with exponentially decaying tails, and these tails could conceivably allow for a nonlocal response to a boundary condition twist; however in such a scenario the exponential tails would necessarily be always bounded from below, and so the resulting AMBL phase would be very different from any known MBL phases. Thus we conclude for a conventional l-bit MBL phase where the entire Hilbert space splits into a product of a complete set of localized l-bits, 
the many-body $T^2$ cannot change with an inserted $\pi$ flux, resulting in an inconsistency with the helical anomaly.

\begin{figure}
	\includegraphics[width=0.48\textwidth]{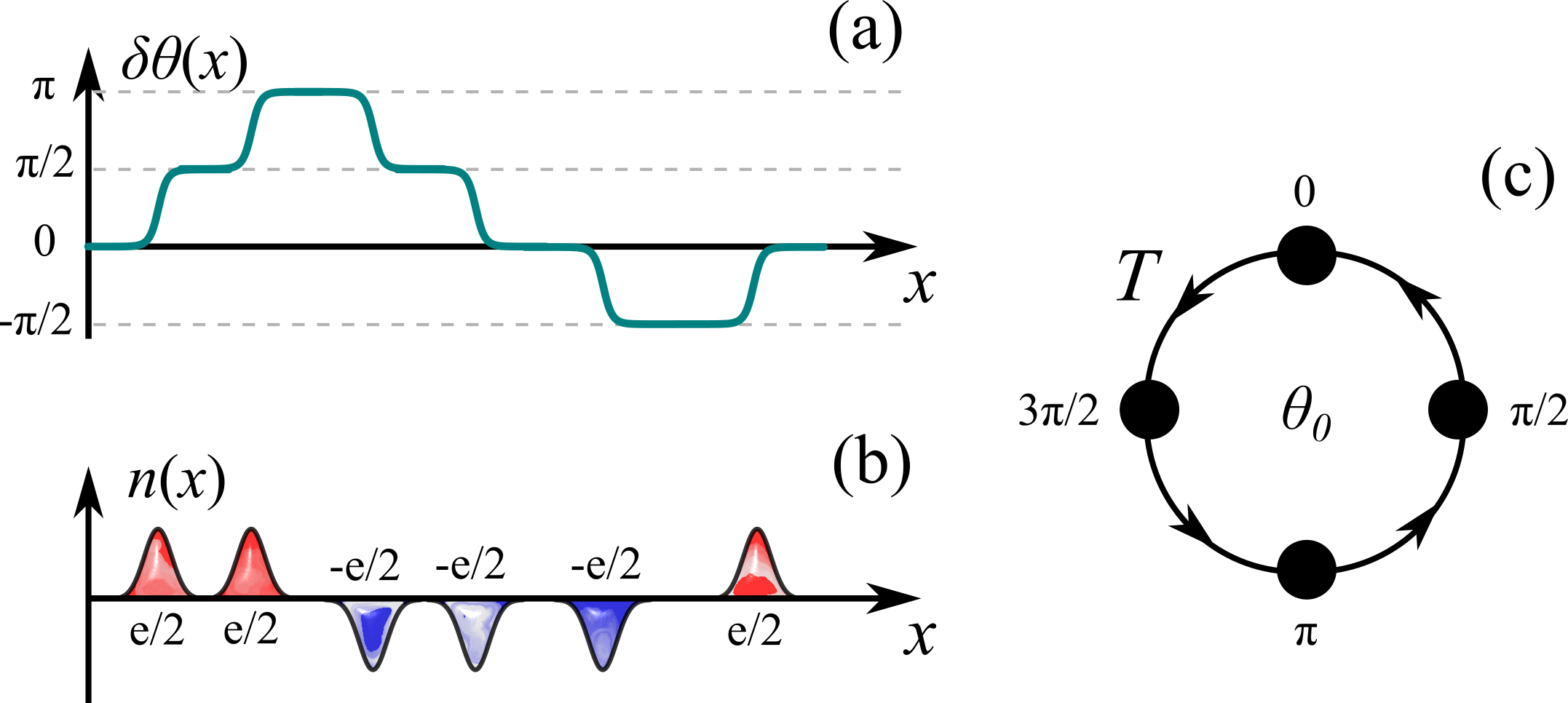}
	\caption{Localized spin glass state on the edge of a 2D TI and its time-reversal transformation. (a) Spatial profile of the local phonon-type  field $\delta\theta(x)$. (b) Density profile of the Luther-Emery fermions, appearing as $T$-breaking-domain walls carrying charge $\pm$e/2. (c) Boson zero mode of the overall shift $\theta_0$; arrows denote its transformation under time-reversal  $T:\ \theta \rightarrow \theta -\pi/2$.
	Panels (a,b) adapted from Ref.~\onlinecite{Chou2018}.}
	\label{fig:dw}
\end{figure}

\subsect{Anomaly manifestation within bosonization}
Relaxing the condition of a complete set of l-bits permits a localized phase to occur in the helical LL. 
We begin in this section by considering a semi-microscopic description, within bosonization, of the Eq.~(\ref{eq:nu}) anomaly in such an AMBL phase. (See the following section for a discussion of how to conceptualize MBL within low energy bosonization in this particular context.) The description is especially useful for a spin glass localized state in a helical LL  (Fig.~\ref{fig:dw}), which has been proposed in early literature\cite{Zhang2006,Moore2006} and recently constructed in detail within re-fermionized bosonization in terms of localized Luther-Emery fermions\cite{Chou2018}, since as discussed above, a fully localized Hilbert space could not realize Eq.~(\ref{eq:nu}).

To define the transformation under time reversal we begin by setting bosonization conventions. 
Define electron right and left moving fields as usual by 
\begin{equation}c_{R,L}(x) \sim e^{i \phi(x)\pm i \theta(x)}. \end{equation}
The electron density $\hat{n}$ and current $\hat{j}$ can be written as $\hat{n}=\partial_x\theta/\pi$ and $\hat{j} = -\partial_t\theta/\pi$.
Time reversal $T$ acts on electrons by taking $c_R\rightarrow  c_L$, $c_L \rightarrow -c_R$ and as always $i \rightarrow -i$, giving the required single particle $T_{\text{single-electron}}^2 = -1$. The corresponding transformation of the $\theta$ and $\phi$ fields under time-reversal $T$ is 
\begin{equation}T:\  \phi \rightarrow -\phi + \pi/2, \ \theta \rightarrow \theta - \pi/2, \ i\rightarrow-i.
\label{eq:Tt}
\end{equation}
At the Luther-Emery point of the Luttinger parameter, the interacting Hamiltonian can be mapped to non-interacting Luther-Emery fermion variables (``refermionized"); away from that parameter point the Luther-Emery fermions interact but can still serve as useful degrees of freedom. The Luther-Emery fermionic fields are conventionally given by
\begin{equation}\Psi_{R,L}(x) \sim e^{i \phi(x)/2 \pm 2 i\theta(x)}.\end{equation} 
The Luther-Emery fermion density corresponds to $\pi/2$ kinks in the $\theta$ field. The action of time reversal on a single Luther-Emery fermion field is given by $\Psi_R\rightarrow e^{i3\pi/4}\Psi_L$, $\Psi_L\rightarrow e^{i3\pi/4}\Psi_R$ (and $i\rightarrow -i$). Importantly, both $\Psi_{L}$ and $\Psi_R$ acquire the same phase $e^{i3\pi/4}$ (since $\phi/2$ appears with the same sign in both); such a phase can be easily absorbed into the fields by an additional factor $e^{i 3 \pi/8}$. 
Indeed regardless of the phase it is easy to see that on each Luther-Emery fermion, the action of time reversal squares to $+1$,
\begin{equation}
T^2_{\text{Luther-Emery fermion}} = +1
\end{equation}
In other words, there are no Kramers doublets within the purely refermionized theory. 
This is consistent since an electron is not just two Luther-Emery  fermions: while two Luther-Emery  fermions correspond to the charge of one electron, they lack the spinful electron's $T_{e}^2=-1$.

As we now show, the bosonization theory captures time-reversal transformations by requiring two additional non-local degrees of freedom, with unusual transformation properties, beyond the Luther-Emery fermions. 
Within bosonization the combination of interactions and disorder leads to the spin-glass state of localized Luther-Emery fermions by producing an operator $\cos 4\theta$. (In contrast $\cos 2\theta$ would break global $T$ symmetry and is forbidden.)
This term stabilizes four states associated with shifting the overall field $\theta_0$ by $[0,\pi/2,\pi,3\pi/2]$. (Depending on the sign of the $\cos 4 \theta $ coefficient these shifts should be interpreted relative to a reference state with $\theta_0=\pi/4$.)  We may refer to them as ``clock states'' and drop the subscript on $\theta_0$ when no ambiguity arises. 
Counting degrees of freedom they are needed to supplement the Luther-Emery fermions, which are $\pi/2$ kinks in $\theta$. This can be understood by analogy to the total spin flip variable in an Ising system, that supplements its description in terms of domain walls; here the number of $\pi/2$ kinks obeys a  constraint that reduces the degrees of freedom by a factor of four rather than two, namely $n_{\text{kinks}} = p \mod 4$, where the particular value of $p$ depends on whether the $\theta$ field obeys periodic boundary conditions ($p=0$) or boundary conditions with a winding number $\pi$ across the system ($p=2$) corresponding to anti-periodic boundary conditions for electrons. We return to this point below when discussing flux insertion, which is a twist of the boundary conditions. 
Note that  there is of course nothing anomalous about the presence of these ``clock states'', which arise quite generally; the anomaly arises based on the time-reversal transformation properties, to which we now turn.

To show how bosonization can consistently capture the anomaly of the helical LL we begin by carefully defining the action of time reversal on the  clock states. 
(Recall that the Luther-Emery fermions have $T^2=+1$.)
According to Eq.~\ref{eq:Tt}, on the clock states time reversal acts as 
$T: \theta \rightarrow \theta - \pi/2$. 
Define states of definite angular momentum on this clock variable: 
\begin{equation}|k{=}k_0\rangle \equiv \sum_{n=0}^{3} e^{i n k_0} |\theta {=}n \pi/2 \rangle.\end{equation}
The label $k$ (periodic in $2\pi$) is meant to be suggestive not only of angular momentum on the clock, but also of physical linear momentum, since $\theta \rightarrow \theta - \pi/2$ is also the action of the global system translation operator $\hat{O}_x$. 
Time reversal acts on the four  $|k\rangle$  states in a simple manner:
$T | k{=} {\pm} \pi/2 \rangle = {\mp} i |k{=}{\mp} \pi/2 \rangle$, while  $T |k{=}0\rangle =  |k{=}0\rangle$ and $T |k{=}\pi\rangle = - |k{=}\pi\rangle$. Equivalently,
\begin{equation}T:\ |k\rangle \rightarrow  e^{-i k}|-k\rangle
\label{eq:Tk}
\end{equation}
in addition to complex conjugation. Hence $T_{MB}^2=-1$ on  the states $ | k{=} {\pm} \pi/2 \rangle $ while $T_{MB}^2=+1$ on the states  $ | k{=}0 \rangle $,  $ | k{=} \pi \rangle $. Since within bosonization there is no other way to set the overall $T_{MB}^2$ of the system, which of these two pairs is a physical state is set by the electron number parity: 
inserting an electron requires switching the sector of the $k$ clock states. (This requirement must be imposed by hand from outside the low energy bosonization theory, as expected for electron insertion.) Within each physical electron parity sector there remain two states of the clock variables. 

Let us contrast the time reversal transformation for an anomalous helical LL with the transformation of a non-anomalous 1D wire. Consider a spinless Luttinger liquid, where microscopic ``electrons" each have $T^2=+1$. Bosonizing these leads to $\theta'$ and $\phi'$ fields that transform as $T: \phi' \rightarrow -\phi'$,  $\theta' \rightarrow \theta'$ (and of course $i\rightarrow-i$), in contrast to Eq.~\ref{eq:Tt}. The invariance of $\theta'$ means that corresponding $k'$ states transform trivially as $T:\ |k'\rangle \rightarrow |-k'\rangle$. 
The case of a spinful 1D Luttinger liquid is distinguished already by having multiple $\theta$ variables corresponding to the two independent spin species. We will return to these two cases after discussing the response to flux insertion  below.

\begin{figure}
	\includegraphics[width=0.35\textwidth]{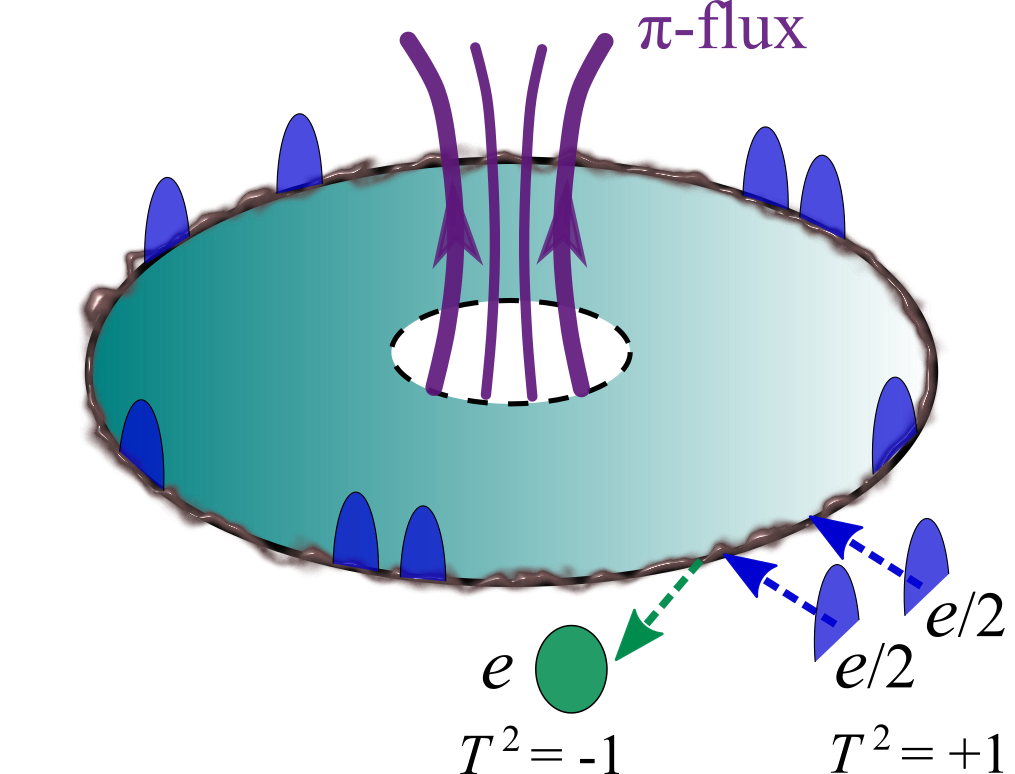}
	\caption{Flux insertion anomaly in the localized state. The antiperiodic boundary conditions for electrons when $\pi$ flux is threaded through the 1D edge can be captured by inserting two Luther-Emery fermions, each carrying charge e/2 and $T^2=+1$. 
	 Tracking many body $T^2$ via the $\theta_0$ ``clock'' variables shows that they are modified by the $\pi$  flux in such a way as to preserve $T^2$; so far, the total charge $n_f$ has changed (say by coupling to a metal) but $T^2$ has not, giving the relative factor $\nu=-1$ in Eq.~(\ref{eq:nu}). 
	Maintaining physical charge neutrality entails formally removing an electron while inserting the e/2 charges; $\theta_0$ changes to account for the electron's $T^2=-1$, again giving the relative factor $\nu=-1$ in Eq.~(\ref{eq:nu}).	}
	\label{fig:flux}
\end{figure}

We now turn to the action of inserting $\pi$ flux enclosed by the edge, or equivalently introducing a $\pi$ twist in the electron's boundary conditions in a periodic 1D helical LL  (Fig.~\ref{fig:flux}). 
This flux insertion does two things in the helical LL. (A) To account for the $\pi$ shift in the boundary conditions for both right and left moving electrons, a consistent representation in terms of $\phi,\theta$ is to leave $\phi$ invariant while modifying $\theta$ boundary conditions so that $\theta(0) = \theta(L) + \pi \mod 2\pi$. This can be done by simply inserting two $\pi/2$ phase-slips/kinks in the $\theta$ field. This corresponds to inserting two Luther-Emery fermions. To keep electric charge fixed this mathematical operation can be supplemented with electron removal/insertion as we elaborate on below. (An alternative representation of flux, that modifies the field $\phi$, would describe states with a finite current that explicitly breaks time reversal and hence are less transparent for the present purpose.)  (B) Since flux insertion also shifts the corresponding linear momentum of states along the edge, and the  $|k\rangle$ states transform like linear momentum eigenstates, the $\pi$ flux insertion also shifts the  $|k\rangle$ states by $\pi$, namely as $|k{=} k_0\rangle \rightarrow  |k{=} k_0 + \pi\rangle$.

Now observe that action (A) changes the electric charge of the system by 1, since the Luther-Emery fermions each carry half an electron's charge. However, since this action (unlike electron insertion) must be describable within the low energy bosonization theory, it cannot (unlike electron insertion) force a change in the clock sector to give a corresponding change in $T^2$. The change in the clock variables so far is due to action (B): but importantly, since it is a $\pi$ shift, though it modifies the states it nevertheless preserves the sign of $T^2$. Thus the flux insertion changes the relation between electronic charge $n_f$ and Kramers parity $T^2$, consistent with the helical anomaly $T^2 = (-1)^{n_f} \nu $ where the $\pi$ flux gives $\nu\rightarrow -1$. Flux insertion within a sector of fixed electric charge corresponds to the operation above together with an electron insertion or removal so as to keep electric charge fixed; but since this electron has $T^2=-1$ (carried within the theory by the respective transformation of the $\theta_0$ clock variables), the  relation between many body $T^2$ and electron number indeed becomes $T^2 = (-1)\times (-1)^{n_f}$, corresponding to $\nu=-1$. This defines the manner in which a charge-neutral Kramers doublet enters (via the 2D TI bulk) into the 1D helical LL, changing $T^2$ relative to charge $n_f$.

Let us again contrast this response to flux insertion to that of 1D non-anomalous spinless and spinful Luttinger Liquids. In the spinless case, both the ``electrons" and the Luther-Emery fermions carry $T^2=+1$ hence flux insertion at fixed charge trivially does nothing. In the spinful case, which can be considered as two copies of a helical LL, the insertion of $\pi$ flux is implemented as a $\pi$ kink in two different $\theta'$ fields, corresponding to four Luther-Emery fermions; fixing the total electric charge involves two electrons insertions/removals, which together are a Kramers singlet $T^2=(-1)(-1)=+1$. No factor $\nu$ can appear. 

The final physical states of the system are combinations of $ |\theta  \rangle$ or  $ |k  \rangle$ states with definite $T^2$ but that break time reversal. 
Listing them by the $T^2$ sector (corresponding to fermion parity times $\pi$-flux parity per Eq.~(\ref{eq:nu})), and in each sector by the two states that are time-reversal pairs, we find (writing in the $k$ and $\theta$ basis respectively),
\begin{align}
T^2= +1: \quad & |k{=}0  \rangle \pm  |k{=}\pi  \rangle 
\\
\text{or}\quad & |\theta{=}0  \rangle +  |\theta{=}\pi  \rangle, \ |\theta{=}\pi/2  \rangle +  |\theta{=}3\pi/2  \rangle ;
\\
T^2=-1: \quad & |k{=}\pi/2  \rangle \pm  |k{=}3\pi/2  \rangle 
\\
\text{or}\quad &
 |\theta{=}0  \rangle -  |\theta{=}\pi  \rangle, \ |\theta{=}\pi/2  \rangle -  |\theta{=}3\pi/2  \rangle .
\end{align}
While superficially these physical states appear to still be cat state superpositions in terms of the clock variable $\theta_0$ (we now restore the $\theta_0$ subscript), that is a misleading consequence of the nonlocal $\theta_0$ representation: since it is only possible to measure $\theta_0 \mod \pi$ with local operators, but not to distinguish $0$ from $\pi$, these  states do not suffer any collapse or decoherence while measuring local operators within a single physical fermion parity sector, and thus are physical (non-cat) states that each belong to a single superselection sector.

\subsect{Anomaly, l-bit and bosonization}

Finally let us comment on the relation between this manifestation of the anomaly in bosonization to the general anomaly restrictions on l-bit MBL phases. 
One way to frame the question is as follows. The anomaly prevents a full lattice product space regularization with onsite action of $T$. 
How, then, can the anomalous system be captured within bosonization?
Bosonization is a low energy theory, but it can sometimes be regularized as a lattice product Hilbert space in terms of some variables. The example relevant here is bosonization in its refermionized form: in the phase where the Luther-Emery fermions are localized, each Luther-Emery fermion can be viewed as defining a Fock space on a lattice site (where sites on the new lattice correspond to some of the bonds on the original microscopic lattice). Within the Luther-Emery fermion sector, this is a regularization with a finite dimensional Hilbert space on each site, and as such in principle could allow l-bit MBL.  

However as the discussion above demonstrates, even as a low energy theory the bosonization must be taken to consist of the Luther-Emery fermion  variables together with the four clock variables. 
One of the two clock ``bits" corresponds to electron parity. 
 The remaining degrees of freedom would form a complete set of l-bits. A state that localizes all remaining degrees of freedom into l-bits is disallowed. However, the remaining clock degree of freedom cannot be localized: it is responsible for the response to time reversal that gives the anomaly. Since time reversal symmetry is preserved on average, considering this clock degree of freedom as an l-bit makes it maximally delocalized across the entire system, since the two symmetry broken states differ by a product of time reversal across all l-bits. The proposed lattice regularization therefore fails in that there remains one degree of freedom which is completely nonlocal relative to the lattice.

We emphasize again that it is not the clock variable itself that is anomalous. Related variables are part of conventional bosonization. Rather, the transformation properties of the clock variables here, and their relation with time reversal and flux insertion, give the anomaly. This distinguishes the present state from states of a 1D system with spontaneous symmetry breaking of non-anomalous time reversal.

MBL has long been known to be allowed to coexist with spin-glass type spontaneous symmetry breaking \cite{Sondhi2013} (SG-SSB). The AMBL with SG-SSB described here is distinct (beyond exhibiting the anomalous responses) in that the SG-SSB is required to occur in every eigenstate as a condition for a fully localized phase to exist.
Interestingly, a ``bubble'' \cite{Schiulaz2016} instability of an MBL locator construction has been used to argue (assuming the instability implies delocalization) that in general the transition into MBL states should occur simultaneously for all eigenstates. Here we find that the anomaly forces the MBL transition to co-occur with SG-SSB simultaneously for all eigenstates  through a mechanism that appears to be quite different from the bubble argument.

\sect{Anomalous MBL phase on the surface of a 3D weak TI and its anomaly manifestation}
\subsect{Construction of 3D weak TI Anomalous-MBL surface}
Here we construct a localized state on a weak TI surface (Fig.~\ref{fig:wti}). Throughout this section we will consider a surface that is not perpendicular to the 3D weak TI stacking vector (the vector associated with its nontrivial $Z_2$ index), i.e.\ we may consider a side surface or any generic surface.

For context let us first recall the noninteracting electron results, namely the mechanism of enforced surface delocalization via symmetry preserving disorder in weak TI surface states with no electron interactions. \cite{Stern2012,FuKaneTopology2012,Moore2012,Akhmerov2014,Mudry2015}
While the clean surface is delocalized (with a dispersion consisting of two Dirac cones), breaking the protecting $Z_2$ translation symmetry can give a fully gapped localized state. 
(Recall that the protecting $Z_2$ symmetry is the translation, mod 2, along the stacking axis of the 2D TI stack; breaking this symmetry corresponds to a dimerization pattern of the stack, which is a $Z_2$ order parameter corresponding to hybridizing odd (or, for the other order parameter domain, even) numbered layers with the layer above, thus leading to a trivial gapped state within each dimerization pattern.) 
The anomaly is then  carried on the surface by defects of the $Z_2$ order parameter, namely domain walls. 
If instead the broken symmetry arises due to disorder such that the net (average) symmetry-breaking order parameter exactly vanishes, then the translation-breaking domains on the surface constitute an ensemble in which each member breaks the symmetry but the ensemble itself is invariant under symmetry transformations. In this case the symmetry is said to be preserved on average and the ensemble describes a system at a transition between the two distinct $Z_2$-translation symmetry-broken states. For the weak TI this surface state, with disordered configurations of the two dimerization patterns and no net dimerization order, also corresponds to the boundary between 2D trivial and topological insulators. Physically, domain walls between the two dimerization patterns host an edge mode of a 2D TI namely a 1D helical LL. Since the symmetry is preserved on average, the network of 1D helical modes is tuned to  its 2D percolation transition, and an extended helical LL spans the surface.

Naively, one might expect that disorder and interactions would lead the surface into such a percolating network of 1D helical modes that would then be localized, per the discussion in Section~\ref{sec:2dti} above, by the very same disorder. 
However that is not the case since the effective disorder experienced by the 1D modes is  correlated due to the network's  spatial optimization. 
In particular each surface helical LL shifts its spatial position to choose to sample optimized regions of the 2D disorder. The generic 1D disorder distribution necessary to localize the 1D helical LL can be avoided when the helical LL is embedded in 2D and preferentially samples a  non-generic distribution. Pinning its position with stronger disorder also increases the density of the domain walls across the system and nucleates many  helical modes arbitrarily close to the spanning mode. The strong disorder limit would then give a dense network of helical modes that interact through these junctions, a setting which is difficult to control theoretically. 

Instead, we construct a localized 2D surface using a double disorder construction, as follows. 
Recall that disorder should be considered just on the surface so as to ensure the bulk gap is preserved. Each disorder realization we consider here  preserves time reversal symmetry exactly.

(A) First we add weak disorder that locally breaks the $Z_2$-translation symmetry that protects the TI (this disorder corresponds to a spatially varying dimerization pattern) but preserves it on average.
Each disorder realization locally breaks the symmetry, but the disorder distribution preserves it exactly.  
 By the Imry-Ma mechanism \cite{Ma1975} (as extended to the marginally relevant case of 2D by Binder \cite{Binder1983}), weak amplitude random ``field'' disorder necessarily gives large domains of the two dimerization patterns, and a sparse configuration of domain walls. 
Since the symmetry is preserved on average the two domains are exactly in the 2D percolation transition, and there must be a domain wall that spans the system. The low energy theory is then a configuration of helical LLs bound to these domain walls including a helical LL that spans the system. We are interested in this spanning helical LL. 

(B) Second we add stronger fully-symmetry-preserving disorder that drives the percolating helical LL to its spin-glass insulating phase \cite{Chou2018} (where time reversal symmetry is locally spontaneously broken but preserved on average). The ``random field" disorder (A) is weak so that the Imry-Ma mechanism's Larkin length  separating domain walls is large, but its presence ensures that the percolating helical LL cannot move around to sample optimized regions of disorder (B), but rather is pinned into generic sampling of the disorder (B).

With only rare regions of LL junctions, whose density is controlled by an independent small parameter, the analytical control of the 1D theory persists to the 2D network on the surface, resulting in the localized surface phase. 
The critical interaction strength needed to drive the surface insulating may in fact be smaller than predicted by the helical LL theory because of the ``clogging effect'' pointed out in Ref.~\onlinecite{Radzihovsky2019}.


\subsect{Nonlinear transport}

\begin{figure}
	\includegraphics[width=0.48\textwidth]{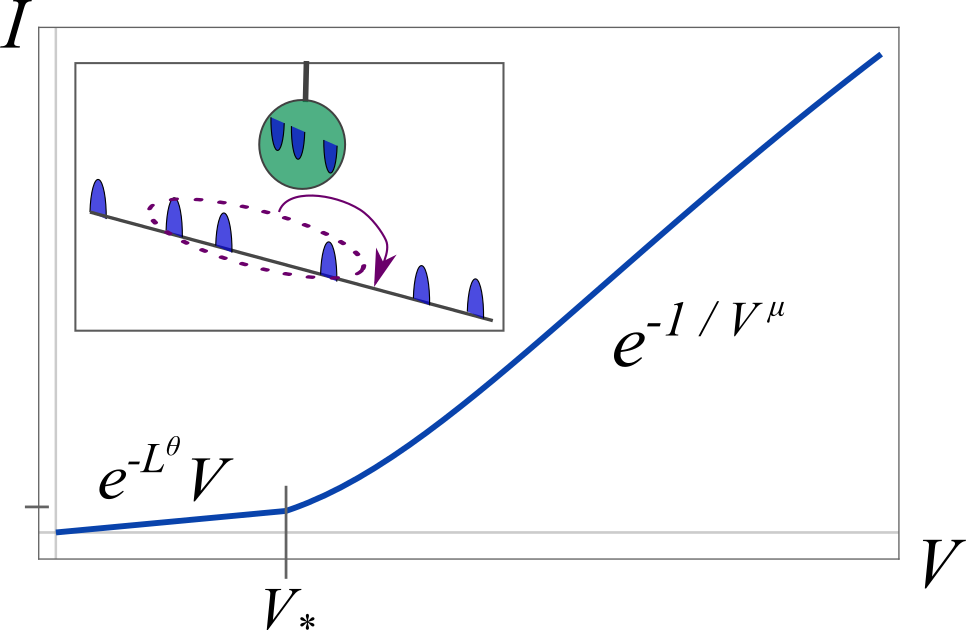}
	\caption{Nonlinear transport and e/2 shot noise. Main panel: crossover in  Eq.~(\ref{eq:IV}) of the nonlinear IV hopping transport from heterogeneous nucleation theory. Inset: illustration of the setup for probing the e/2 quantized shot noise. Transport proceeds via thermally activated or quantum tunneling events wherein a droplet of localized e/2 charges rearranges and shifts in an applied voltage. An electrostatic probe such as a Kelvin probe (green disk) captures the electric field lines of the carriers via their image charge. The resulting current shot noise shows e/2 quantization.}
	\label{fig:iv}
\end{figure}

There are two distinct ways to manifest helical modes in the 3D weak TI, as illustrate in Fig.~\ref{fig:wti}: through its surface state or through a defect inserted in its bulk. 
For noninteracting electrons, the signatures of the topological state are a metallic surface, and an insulating bulk that shows unusual response to a ``probe'' dislocation defect: recall  \cite{Vishwanath2009} that a helical LL is bound to a dislocation when $\vec G_0/2 \cdot \vec{b} = \pi$, where $\vec G_0$ is the reciprocal lattice vector corresponding to the weak TI stacking   and $\vec b$ is the dislocation's Burgers vector (a lattice vector denoting the shift of all atoms on the 2D slice bounded by the dislocation). 
Now consider the case of strong electron interactions. The surface may show the new AMBL phase constructed above, which  has gapless excitations localized to small regions as in the  Bose glass phase \cite{Fisher1989}, but not metallic. Similarly a helical LL bound to a dislocation in the bulk can show its own AMBL phase. How can this helical LL be observed if it is AMBL?

In this section we discuss the nonlinear hopping transport exhibited by the AMBL helical LL. This is  conventional (non-anomalous) phenomenology for glassy systems, here also extended to SG-SSB MBL; direct experimental probes of the anomaly in AMBL will be discussed in the following section. 

Before we begin, let us note an interesting feature of helical LLs arising in 3D weak TIs: they can naturally be finite-sized 1D systems even if the 3D system is infinite. On the surface in its AMBL construction above, the helical LLs away from the percolating mode are finite-sized. In the bulk, the anomaly is locally probed by a finite edge dislocation, e.g.\ an insertion of a finite 2D TI slice perpendicular to the surface such that the helical LL passes through the bulk in a ``U'' shape with finite length. To study the phenomenology of these would-be anomaly probes we are thus led to consider transport both in the thermodynamic limit and in finite sized 1D systems. 

Transport in the AMBL helical LL shows a sub-linear current response to an applied voltage, which we now proceed to compute (Fig.~\ref{fig:iv}). To discuss an electric field applied to a localized system, we assume that the electron Hilbert space connects to another Hilbert space at high energies, so as to avoid the trivial effect where  a sufficiently large electric field trivially leads to enhanced localization  with level spacing of the Bloch oscillation frequency. This can be resolved by e.g.\ coupling to a phonon bath, though the coupling should be weak: as always with MBL, to discuss states at finite temperature or finite energy density we assume that the system is effectively sufficiently closed at the time scales probed so that thermalization is avoided. With these assumptions we may consider both finite temperature $T>0$ and zero temperature $T=0$ transport, beginning with $T>0$. 

Recall the results of heterogeneous nucleation theory \cite{Fisher1967,Ambegaokar1967,Halperin1970,Huse1986,Huse1991}, which describes creep motion via collective rearrangements in many systems pinned by impurities and subject to an external perturbation, including pinned charge density waves and superconductor vortex lattices. 
With the external perturbation $V$, the system finds itself in a metastable state that is separated from a favored lower energy state by large energy barriers $U$. The energy barrier $U$ for an advantageous rearrangement is a growing function of the size $L$ of the rearranged region, $U(L) \sim L^\theta$ with exponent $\theta>0$. (Henceforth we use the symbol $\theta$ to denote this exponent.)
Given the external perturbation $V$, the nucleation scaling theory predicts an optimal rearrangement involving a droplet of size $L_\star(V) \sim V^{-\mu/\theta}$, with another exponent $\mu>0$. 
A small external perturbation yields an optimal rearrangement of a large droplet, where for example this rearrangement may entail a de-pinning and minimal shift of a charge density wave within the $L_\star$ droplet region. Relaxation from the metastable state then occurs by nucleating such droplets with energy $U(L_\star(V)) \sim V^{-\mu}$ at a frequency given by their Boltzmann weight $e^{-U/T}$. 

Here we apply the heterogeneous nucleation theory to the present case, the current response $I$ to an applied voltage $V$ of a pinned 1D array of Luther-Emery fermions. At thermal equilibrium, $T>0$ in a 1D system would produce finite-energy barriers for large rearrangements i.e.\ $\theta<0$, so that the  putative metastable state is actually unstable as must be the case for $T>0$ 1D ordered states in equilibrium, resulting in  linear $IV$ when $V$ is small compared to temperature. However here the spin-glass-ordered AMBL state avoids thermal equilibrium and remains stable at finite energy density, so we expect it will be described by heterogeneous nucleation theory with $\theta>0$, giving nonlinear IV even at finite energy density. Such finite-energy-density nonlinear IV in 1D serves as a fingerprint of MBL transport.

  At zero temperature the conventional nucleation theory predicts rearrangements determined by their spacetime action cost $S(V) \sim V^{-\mu}$, where the $T=0$ exponent $\mu$ can differ from the finite temperature one (and is likely larger corresponding to a larger exponent $\theta$ for the cost of spacetime rearrangements); and with their thermal excitation frequency replaced by a rate of quantum tunneling as $e^{-U/T} \rightarrow e^{-S/\hbar}$. 
  
 In both cases, the nonlinear IV that occurs for sufficiently large applied voltage or system size is thus cut off into linear exponentially-suppressed conductivity when the system size becomes smaller than the optimal rearrangement size $L_\star(V)$, or equivalently
\begin{equation} \label{eq:IV}
I \sim
\begin{cases}
e^{-1/V^{\mu}} & V > V_\star(L) \sim L^{-\theta/\mu} \\
 \sigma V &  V < V_\star(L) \sim L^{-\theta/\mu}
\end{cases} \end{equation}
where $\sigma \sim e^{-L^\theta}$. This form is shown in Fig.~\ref{fig:iv}.

It is instructive to compare these expressions with recent transport calculations for weakly interacting strongly disordered 2D TI edges in the large nonlinear voltage regime. \cite{Rudner2019}  In that case the linear response conductivity $I \sim V$ crosses over, at large currents or voltages, to the form $I \sim V^{1/3}$. However the mechanism in that case is due to the current explicitly breaking time reversal, and requires much larger voltages than the nonlinear form in Eq.~(\ref{eq:IV}). 

\subsect{Probing the anomaly via e/2 shot noise}
\label{sec:halfcharge}
To probe the anomaly, on a 2D TI edge or on the 3D weak TI surface, even when it enters an AMBL phase, we consider the anomaly feature of $\pm$e/2 electric charges on the point-like domain wall between opposite time-reversal-breaking domains. 
If a weak TI surface is in the AMBL phase, such  e/2 charges will appear at finite density across the  helical network spanning the surface, both in the percolating helical LL and in domain boundaries of finite domains.

The e/2 charges could be directly observed electrostatically using a local probe that can sense the local charge distribution. 
Local electrostatic sensing is routinely performed using Kelvin probes, including for topological states\cite{GossLevi1998}. Since the e/2 charges are confined to the helical LL, it is important that the probe couple to their electric field lines without requiring direct charge hopping. 
Such electrostatic coupling is related to capacitive coupling, albeit here the coupling is to an insulator; rather it can be better interpreted via image charges forming on the metallic probe, corresponding to the nearby charge on the insulator.
However while such a static charge measurement is in principle possible, distinguishing a localized charge e/2 from an electron would be complicated by the local static polarizability of the insulator.
The dielectric polarizability could possibly be measured independently by applying a strong uniform magnetic field so as to replace the AMBL phase with a ferromagnet of localized electrons, which (one can hope) may have a similar dielectric constant which could then be used for calibration.

Instead, a likely better measurement of the e/2 charges can be performed using shot noise  associated with the hopping transport discussed above, as follows. First recall the setting for conventional current shot noise measured across a tunnel junction. The current noise across the junction, $S_I$, is simply related to the average current times the carrier charge, $S_I = 2 e I$, at essentially all frequencies smaller than the inverse junction crossing time of a single electron. The measuring probe need not (and should not) resolve the time window of a single charge  hopping event; instead the charge quantization is extracted from the statistics of many hopping events averaged over a large time window. 

Similar considerations apply for shot noise in the present case. The measuring probe cannot directly couple to current of the e/2 charges (which only arise inside the helical LL), but instead can couple using a Kelvin-probe-type electrostatic measurement as discussed above  (Fig.~\ref{fig:iv}). It could be an STM-like scanning tip or a fixed electrode on a substrate.
A probe with infinite capacitance would show instantaneous image charges. Though the fraction of electric field lines and thus of the image charge recorded by the probe would depend on the local geometry, in principle such a ``charge efficiency factor'' can be calibrated. 
The main distinction is in the possible introduction of an additional time scale $\tau$ arising from the kinematics of the hopping transport, as follows. The hopping transport occurs via local depinning of a cluster of domains and e/2 charged domain walls, which then hop to another position. The movement of the individual e/2 charges can be detected by the probe. Since the resulting shot noise arises from the induced electrostatic charge on the probe, rather than a current, it is modified from the current shot noise by a factor $\tau$ with dimensions of time, related to the time spent performing each hop. The time $\tau$ will depend e.g.\ on the size of the hopping cluster, though such detailed kinematics are not predicted by the stochastic nucleation scaling theory. However  the dependence on kinematic details may be eliminated in an appropriate regime: when the Luther-Emery fermions are relatively sparse, so that the probe can measure the charge across a number of lattice sites that is still smaller than typical Luther-Emery fermion separation, then the time $\tau$ is effectively averaged across multiple hops so that only an average velocity needs to be known to extract the current shot noise. Leaving details of this proposed experiment aside, it is clear that it is at least in principle possible to directly probe the anomaly even in the AMBL state.

In addition to this specific measurement of the e/2 charges, it is worth noting that the AMBL localized state is distinguished from electron spin glasses by the fractionalized nature of the localized $\pm$e/2 Luther-Emery fermions. Their creation operators are related to electron creation operators in a nonlocal manner. This can be formulated via entanglement and information scrambling as follows. \cite{Yao2017} Consider an electron, entangled with a measurement device, for example in a singlet state with a reference electron, tunneling into the localized state. While in a conventional spin glass, the electron and its information will remain localized, in the AMBL spin glass the information encoded is quickly scrambled across the entire system, as the entanglement with the reference electron delocalizes. The quantum information conveyed by the tunneled electron becomes a nonlocal qubit, protected by the localized boundary anomaly.

\sect{Conclusion and outlook}
In this work we studied how localization due to strong disorder and interactions can coexist with the boundary anomaly of a helical LL. The resulting AMBL phase shows interesting features both at the level of the Hilbert space and for direct experimental observables, that can distinguish it from conventional MBL states. The present study is applied to 2D TIs and 3D weak TIs. 

In contrast, on the surface of 3D strong TIs locally-symmetry-breaking disorder leads to a network of chiral LLs non-localizable even with strong interactions, suggesting a qualitative distinction in allowed phases manifested by combining strong interactions and statistically-symmetric disorder.
We find that symmetries that are preserved statistically (``on average''), though locally broken by disorder, play a key role even in the presence of arbitrarily strong electron interactions. 
They can preserve the symmetry-protected anomaly albeit with potentially modified consequences: here we found that a localized phase can arise, with a description in terms of electrons that satisfies certain constraints, thereby exhibiting the required  anomaly responses to external probes.

Looking beyond electronic TIs more broadly at interacting symmetry-protected topological phases, it is intriguing to consider a more general viewpoint on boundary anomalies. As has been known for some time \cite{Bonderson2016}, a boundary anomaly can sometimes be viewed in terms of a Lieb-Schultz-Mattis (LSM) type theorem \cite{Mattis1961, Oshikawa2000}, which restricts the range of possible allowed phases under arbitrary interactions, if the lattice unit cell obeys certain conditions. For instance an array of spin-1 gapped Haldane chains has as its boundary a  spin-1/2 lattice which obeys an LSM theorem. More recently it has been argued \cite{randommagnets} that spin-1/2 magnets should satisfy disordered-LSM-type restrictions even though the disorder breaks the translation symmetry that defines the unit cell, as long as translation symmetry is preserved on average. In 1D systems a robust argument is available  \cite{randommagnets} and the disordered-LSM restriction can enforce  nonlocal features such as at least $1/r^2$ long range average spin correlations. This is the ``minimal anomaly manifestation'' required for arbitrary interactions by the disordered-LSM restriction (which, interestingly, is realized in the random-singlet fixed point \cite{Fisher1994}). For the helical LL, we find that the minimal anomaly manifestation is consistent with AMBL, a type of localization. This applies to 3D weak TIs but is not the case for 3D strong TIs. Understanding how to extract any particular minimal anomaly manifestation, in the presence of strong interactions and disorder, is an interesting question for framing future studies.

\

\textit{Acknowledgements.} We thank Jason Iaconis, Victor Gurarie, Max Metlitski, Charles Rogers and John Price for helpful discussions, and especially Michael Hermele for in depth discussions of ideas and encouragement throughout this project.

This work is supported in part by  the DARPA DRINQS program (I.K., Y.-Z.C. and R.M.N.), in part by a National Research Council Fellowship through the National Institute of Standards and Technology (I.K.), in part by a Simons Investigator award from the Simons Foundation (L.R., Y.-Z.C.), and in part by the Alfred P. Sloan Foundation through a Sloan Research Fellowship (R.M.N.).

\textit{Note added.} During completion of this work, Ref.~\onlinecite{Metlitski2019} was posted   constructing a 1D lattice model for the boundary of a 2D time-reversal-protected fermionic topological insulator, using local spins and fermions with unusual (non-onsite) symmetry transformation. (See also a related preprint Ref \onlinecite{Alicea2019}.) 
A logical corollary that follows from the construction provides an AMBL state with anomalous responses and locally broken time-reversal symmetry. We note that though Ref \onlinecite{Metlitski2019} also discusses a particular bosonization theory on this Hilbert space, those bosonization fields do not seem to be related via any local transformation to the fields arising from bosonization of helical electrons, and in particular the anomaly manifestation of inserting $\pi$ flux (changing electronic boundary conditions) is distinct as well.

\bibliography{TIrefs}

\begin{thebibliography}{57}%
\makeatletter
\providecommand \@ifxundefined [1]{%
 \@ifx{#1\undefined}
}%
\providecommand \@ifnum [1]{%
 \ifnum #1\expandafter \@firstoftwo
 \else \expandafter \@secondoftwo
 \fi
}%
\providecommand \@ifx [1]{%
 \ifx #1\expandafter \@firstoftwo
 \else \expandafter \@secondoftwo
 \fi
}%
\providecommand \natexlab [1]{#1}%
\providecommand \enquote  [1]{``#1''}%
\providecommand \bibnamefont  [1]{#1}%
\providecommand \bibfnamefont [1]{#1}%
\providecommand \citenamefont [1]{#1}%
\providecommand \href@noop [0]{\@secondoftwo}%
\providecommand \href [0]{\begingroup \@sanitize@url \@href}%
\providecommand \@href[1]{\@@startlink{#1}\@@href}%
\providecommand \@@href[1]{\endgroup#1\@@endlink}%
\providecommand \@sanitize@url [0]{\catcode `\\12\catcode `\$12\catcode
  `\&12\catcode `\#12\catcode `\^12\catcode `\_12\catcode `\%12\relax}%
\providecommand \@@startlink[1]{}%
\providecommand \@@endlink[0]{}%
\providecommand \url  [0]{\begingroup\@sanitize@url \@url }%
\providecommand \@url [1]{\endgroup\@href {#1}{\urlprefix }}%
\providecommand \urlprefix  [0]{URL }%
\providecommand \Eprint [0]{\href }%
\providecommand \doibase [0]{http://dx.doi.org/}%
\providecommand \selectlanguage [0]{\@gobble}%
\providecommand \bibinfo  [0]{\@secondoftwo}%
\providecommand \bibfield  [0]{\@secondoftwo}%
\providecommand \translation [1]{[#1]}%
\providecommand \BibitemOpen [0]{}%
\providecommand \bibitemStop [0]{}%
\providecommand \bibitemNoStop [0]{.\EOS\space}%
\providecommand \EOS [0]{\spacefactor3000\relax}%
\providecommand \BibitemShut  [1]{\csname bibitem#1\endcsname}%
\let\auto@bib@innerbib\@empty
\bibitem [{\citenamefont {Kane}\ and\ \citenamefont
  {Mele}(2005{\natexlab{a}})}]{Mele2005}%
  \BibitemOpen
  \bibfield  {author} {\bibinfo {author} {\bibfnamefont {C.~L.}\ \bibnamefont
  {Kane}}\ and\ \bibinfo {author} {\bibfnamefont {E.~J.}\ \bibnamefont
  {Mele}},\ }\bibfield  {title} {\enquote {\bibinfo {title} {${Z}_{2}$
  topological order and the quantum spin hall effect},}\ }\href {\doibase
  10.1103/PhysRevLett.95.146802} {\bibfield  {journal} {\bibinfo  {journal}
  {Phys. Rev. Lett.}\ }\textbf {\bibinfo {volume} {95}},\ \bibinfo {pages}
  {146802} (\bibinfo {year} {2005}{\natexlab{a}})}\BibitemShut {NoStop}%
\bibitem [{\citenamefont {Kane}\ and\ \citenamefont
  {Mele}(2005{\natexlab{b}})}]{Mele2005a}%
  \BibitemOpen
  \bibfield  {author} {\bibinfo {author} {\bibfnamefont {C.~L.}\ \bibnamefont
  {Kane}}\ and\ \bibinfo {author} {\bibfnamefont {E.~J.}\ \bibnamefont
  {Mele}},\ }\bibfield  {title} {\enquote {\bibinfo {title} {Quantum spin hall
  effect in graphene},}\ }\href {\doibase 10.1103/PhysRevLett.95.226801}
  {\bibfield  {journal} {\bibinfo  {journal} {Phys. Rev. Lett.}\ }\textbf
  {\bibinfo {volume} {95}},\ \bibinfo {pages} {226801} (\bibinfo {year}
  {2005}{\natexlab{b}})}\BibitemShut {NoStop}%
\bibitem [{\citenamefont {Fu}\ \emph {et~al.}(2007)\citenamefont {Fu},
  \citenamefont {Kane},\ and\ \citenamefont {Mele}}]{Mele2007}%
  \BibitemOpen
  \bibfield  {author} {\bibinfo {author} {\bibfnamefont {Liang}\ \bibnamefont
  {Fu}}, \bibinfo {author} {\bibfnamefont {C.~L.}\ \bibnamefont {Kane}}, \ and\
  \bibinfo {author} {\bibfnamefont {E.~J.}\ \bibnamefont {Mele}},\ }\bibfield
  {title} {\enquote {\bibinfo {title} {Topological insulators in three
  dimensions},}\ }\href {\doibase 10.1103/PhysRevLett.98.106803} {\bibfield
  {journal} {\bibinfo  {journal} {Phys. Rev. Lett.}\ }\textbf {\bibinfo
  {volume} {98}},\ \bibinfo {pages} {106803} (\bibinfo {year}
  {2007})}\BibitemShut {NoStop}%
\bibitem [{\citenamefont {Moore}\ and\ \citenamefont
  {Balents}(2007)}]{Balents2007}%
  \BibitemOpen
  \bibfield  {author} {\bibinfo {author} {\bibfnamefont {J.~E.}\ \bibnamefont
  {Moore}}\ and\ \bibinfo {author} {\bibfnamefont {L.}~\bibnamefont
  {Balents}},\ }\bibfield  {title} {\enquote {\bibinfo {title} {Topological
  invariants of time-reversal-invariant band structures},}\ }\href {\doibase
  10.1103/PhysRevB.75.121306} {\bibfield  {journal} {\bibinfo  {journal} {Phys.
  Rev. B}\ }\textbf {\bibinfo {volume} {75}},\ \bibinfo {pages} {121306}
  (\bibinfo {year} {2007})}\BibitemShut {NoStop}%
\bibitem [{\citenamefont {Roy}(2009)}]{Roy2009}%
  \BibitemOpen
  \bibfield  {author} {\bibinfo {author} {\bibfnamefont {Rahul}\ \bibnamefont
  {Roy}},\ }\bibfield  {title} {\enquote {\bibinfo {title} {Topological phases
  and the quantum spin hall effect in three dimensions},}\ }\href {\doibase
  10.1103/PhysRevB.79.195322} {\bibfield  {journal} {\bibinfo  {journal} {Phys.
  Rev. B}\ }\textbf {\bibinfo {volume} {79}},\ \bibinfo {pages} {195322}
  (\bibinfo {year} {2009})}\BibitemShut {NoStop}%
\bibitem [{\citenamefont {Hasan}\ and\ \citenamefont {Kane}(2010)}]{Kane2010}%
  \BibitemOpen
  \bibfield  {author} {\bibinfo {author} {\bibfnamefont {M.~Z.}\ \bibnamefont
  {Hasan}}\ and\ \bibinfo {author} {\bibfnamefont {C.~L.}\ \bibnamefont
  {Kane}},\ }\bibfield  {title} {\enquote {\bibinfo {title} {Colloquium:
  Topological insulators},}\ }\href {\doibase 10.1103/RevModPhys.82.3045}
  {\bibfield  {journal} {\bibinfo  {journal} {Rev. Mod. Phys.}\ }\textbf
  {\bibinfo {volume} {82}},\ \bibinfo {pages} {3045--3067} (\bibinfo {year}
  {2010})}\BibitemShut {NoStop}%
\bibitem [{\citenamefont {Qi}\ and\ \citenamefont {Zhang}(2011)}]{Zhang2011}%
  \BibitemOpen
  \bibfield  {author} {\bibinfo {author} {\bibfnamefont {Xiao-Liang}\
  \bibnamefont {Qi}}\ and\ \bibinfo {author} {\bibfnamefont {Shou-Cheng}\
  \bibnamefont {Zhang}},\ }\bibfield  {title} {\enquote {\bibinfo {title}
  {Topological insulators and superconductors},}\ }\href {\doibase
  10.1103/RevModPhys.83.1057} {\bibfield  {journal} {\bibinfo  {journal} {Rev.
  Mod. Phys.}\ }\textbf {\bibinfo {volume} {83}},\ \bibinfo {pages}
  {1057--1110} (\bibinfo {year} {2011})}\BibitemShut {NoStop}%
\bibitem [{\citenamefont {Ringel}\ \emph {et~al.}(2012)\citenamefont {Ringel},
  \citenamefont {Kraus},\ and\ \citenamefont {Stern}}]{Stern2012}%
  \BibitemOpen
  \bibfield  {author} {\bibinfo {author} {\bibfnamefont {Zohar}\ \bibnamefont
  {Ringel}}, \bibinfo {author} {\bibfnamefont {Yaacov~E.}\ \bibnamefont
  {Kraus}}, \ and\ \bibinfo {author} {\bibfnamefont {Ady}\ \bibnamefont
  {Stern}},\ }\bibfield  {title} {\enquote {\bibinfo {title} {Strong side of
  weak topological insulators},}\ }\href {\doibase 10.1103/PhysRevB.86.045102}
  {\bibfield  {journal} {\bibinfo  {journal} {Phys. Rev. B}\ }\textbf {\bibinfo
  {volume} {86}},\ \bibinfo {pages} {045102} (\bibinfo {year}
  {2012})}\BibitemShut {NoStop}%
\bibitem [{\citenamefont {Fu}\ and\ \citenamefont
  {Kane}(2012)}]{FuKaneTopology2012}%
  \BibitemOpen
  \bibfield  {author} {\bibinfo {author} {\bibfnamefont {Liang}\ \bibnamefont
  {Fu}}\ and\ \bibinfo {author} {\bibfnamefont {C.~L.}\ \bibnamefont {Kane}},\
  }\bibfield  {title} {\enquote {\bibinfo {title} {Topology, delocalization via
  average symmetry and the symplectic anderson transition},}\ }\href {\doibase
  10.1103/PhysRevLett.109.246605} {\bibfield  {journal} {\bibinfo  {journal}
  {Phys. Rev. Lett.}\ }\textbf {\bibinfo {volume} {109}},\ \bibinfo {pages}
  {246605} (\bibinfo {year} {2012})}\BibitemShut {NoStop}%
\bibitem [{\citenamefont {Mong}\ \emph {et~al.}(2012)\citenamefont {Mong},
  \citenamefont {Bardarson},\ and\ \citenamefont {Moore}}]{Moore2012}%
  \BibitemOpen
  \bibfield  {author} {\bibinfo {author} {\bibfnamefont {Roger S.~K.}\
  \bibnamefont {Mong}}, \bibinfo {author} {\bibfnamefont {Jens~H.}\
  \bibnamefont {Bardarson}}, \ and\ \bibinfo {author} {\bibfnamefont {Joel~E.}\
  \bibnamefont {Moore}},\ }\bibfield  {title} {\enquote {\bibinfo {title}
  {Quantum transport and two-parameter scaling at the surface of a weak
  topological insulator},}\ }\href {\doibase 10.1103/PhysRevLett.108.076804}
  {\bibfield  {journal} {\bibinfo  {journal} {Phys. Rev. Lett.}\ }\textbf
  {\bibinfo {volume} {108}},\ \bibinfo {pages} {076804} (\bibinfo {year}
  {2012})}\BibitemShut {NoStop}%
\bibitem [{\citenamefont {Fulga}\ \emph {et~al.}(2014)\citenamefont {Fulga},
  \citenamefont {van Heck}, \citenamefont {Edge},\ and\ \citenamefont
  {Akhmerov}}]{Akhmerov2014}%
  \BibitemOpen
  \bibfield  {author} {\bibinfo {author} {\bibfnamefont {I.~C.}\ \bibnamefont
  {Fulga}}, \bibinfo {author} {\bibfnamefont {B.}~\bibnamefont {van Heck}},
  \bibinfo {author} {\bibfnamefont {J.~M.}\ \bibnamefont {Edge}}, \ and\
  \bibinfo {author} {\bibfnamefont {A.~R.}\ \bibnamefont {Akhmerov}},\
  }\bibfield  {title} {\enquote {\bibinfo {title} {Statistical topological
  insulators},}\ }\href {\doibase 10.1103/PhysRevB.89.155424} {\bibfield
  {journal} {\bibinfo  {journal} {Phys. Rev. B}\ }\textbf {\bibinfo {volume}
  {89}},\ \bibinfo {pages} {155424} (\bibinfo {year} {2014})}\BibitemShut
  {NoStop}%
\bibitem [{\citenamefont {Morimoto}\ \emph {et~al.}(2015)\citenamefont
  {Morimoto}, \citenamefont {Furusaki},\ and\ \citenamefont
  {Mudry}}]{Mudry2015}%
  \BibitemOpen
  \bibfield  {author} {\bibinfo {author} {\bibfnamefont {Takahiro}\
  \bibnamefont {Morimoto}}, \bibinfo {author} {\bibfnamefont {Akira}\
  \bibnamefont {Furusaki}}, \ and\ \bibinfo {author} {\bibfnamefont
  {Christopher}\ \bibnamefont {Mudry}},\ }\bibfield  {title} {\enquote
  {\bibinfo {title} {Anderson localization and the topology of classifying
  spaces},}\ }\href {\doibase 10.1103/PhysRevB.91.235111} {\bibfield  {journal}
  {\bibinfo  {journal} {Phys. Rev. B}\ }\textbf {\bibinfo {volume} {91}},\
  \bibinfo {pages} {235111} (\bibinfo {year} {2015})}\BibitemShut {NoStop}%
\bibitem [{\citenamefont {Chou}\ \emph {et~al.}(2018)\citenamefont {Chou},
  \citenamefont {Nandkishore},\ and\ \citenamefont {Radzihovsky}}]{Chou2018}%
  \BibitemOpen
  \bibfield  {author} {\bibinfo {author} {\bibfnamefont {Yang-Zhi}\
  \bibnamefont {Chou}}, \bibinfo {author} {\bibfnamefont {Rahul~M.}\
  \bibnamefont {Nandkishore}}, \ and\ \bibinfo {author} {\bibfnamefont {Leo}\
  \bibnamefont {Radzihovsky}},\ }\bibfield  {title} {\enquote {\bibinfo {title}
  {Gapless insulating edges of dirty interacting topological insulators},}\
  }\href {\doibase 10.1103/PhysRevB.98.054205} {\bibfield  {journal} {\bibinfo
  {journal} {Phys. Rev. B}\ }\textbf {\bibinfo {volume} {98}},\ \bibinfo
  {pages} {054205} (\bibinfo {year} {2018})}\BibitemShut {NoStop}%
\bibitem [{\citenamefont {Wang}\ \emph {et~al.}(2013)\citenamefont {Wang},
  \citenamefont {Potter},\ and\ \citenamefont {Senthil}}]{Senthil2013}%
  \BibitemOpen
  \bibfield  {author} {\bibinfo {author} {\bibfnamefont {Chong}\ \bibnamefont
  {Wang}}, \bibinfo {author} {\bibfnamefont {Andrew~C.}\ \bibnamefont
  {Potter}}, \ and\ \bibinfo {author} {\bibfnamefont {T.}~\bibnamefont
  {Senthil}},\ }\bibfield  {title} {\enquote {\bibinfo {title} {Gapped symmetry
  preserving surface state for the electron topological insulator},}\ }\href
  {\doibase 10.1103/PhysRevB.88.115137} {\bibfield  {journal} {\bibinfo
  {journal} {Phys. Rev. B}\ }\textbf {\bibinfo {volume} {88}},\ \bibinfo
  {pages} {115137} (\bibinfo {year} {2013})}\BibitemShut {NoStop}%
\bibitem [{\citenamefont {Chen}\ \emph {et~al.}(2014)\citenamefont {Chen},
  \citenamefont {Fidkowski},\ and\ \citenamefont
  {Vishwanath}}]{Vishwanath2014b}%
  \BibitemOpen
  \bibfield  {author} {\bibinfo {author} {\bibfnamefont {Xie}\ \bibnamefont
  {Chen}}, \bibinfo {author} {\bibfnamefont {Lukasz}\ \bibnamefont
  {Fidkowski}}, \ and\ \bibinfo {author} {\bibfnamefont {Ashvin}\ \bibnamefont
  {Vishwanath}},\ }\bibfield  {title} {\enquote {\bibinfo {title} {Symmetry
  enforced non-abelian topological order at the surface of a topological
  insulator},}\ }\href {\doibase 10.1103/PhysRevB.89.165132} {\bibfield
  {journal} {\bibinfo  {journal} {Phys. Rev. B}\ }\textbf {\bibinfo {volume}
  {89}},\ \bibinfo {pages} {165132} (\bibinfo {year} {2014})}\BibitemShut
  {NoStop}%
\bibitem [{\citenamefont {Bonderson}\ \emph {et~al.}(2013)\citenamefont
  {Bonderson}, \citenamefont {Nayak},\ and\ \citenamefont {Qi}}]{Qi2013}%
  \BibitemOpen
  \bibfield  {author} {\bibinfo {author} {\bibfnamefont {Parsa}\ \bibnamefont
  {Bonderson}}, \bibinfo {author} {\bibfnamefont {Chetan}\ \bibnamefont
  {Nayak}}, \ and\ \bibinfo {author} {\bibfnamefont {Xiao-Liang}\ \bibnamefont
  {Qi}},\ }\bibfield  {title} {\enquote {\bibinfo {title} {A time-reversal
  invariant topological phase at the surface of a 3d topological insulator},}\
  }\href@noop {} {\bibfield  {journal} {\bibinfo  {journal} {Journal of
  Statistical Mechanics: Theory and Experiment}\ }\textbf {\bibinfo {volume}
  {2013}},\ \bibinfo {pages} {P09016} (\bibinfo {year} {2013})}\BibitemShut
  {NoStop}%
\bibitem [{\citenamefont {Ran}\ \emph {et~al.}(2009)\citenamefont {Ran},
  \citenamefont {Zhang},\ and\ \citenamefont {Vishwanath}}]{Vishwanath2009}%
  \BibitemOpen
  \bibfield  {author} {\bibinfo {author} {\bibfnamefont {Ying}\ \bibnamefont
  {Ran}}, \bibinfo {author} {\bibfnamefont {Yi}~\bibnamefont {Zhang}}, \ and\
  \bibinfo {author} {\bibfnamefont {Ashvin}\ \bibnamefont {Vishwanath}},\
  }\bibfield  {title} {\enquote {\bibinfo {title} {One-dimensional
  topologically protected modes in topological insulators with lattice
  dislocations},}\ }\href {https://doi.org/10.1038/nphys1220} {\bibfield
  {journal} {\bibinfo  {journal} {Nature Physics}\ }\textbf {\bibinfo {volume}
  {5}},\ \bibinfo {pages} {298} (\bibinfo {year} {2009})}\BibitemShut {NoStop}%
\bibitem [{\citenamefont {Schnyder}\ \emph {et~al.}(2008)\citenamefont
  {Schnyder}, \citenamefont {Ryu}, \citenamefont {Furusaki},\ and\
  \citenamefont {Ludwig}}]{Ludwig2008}%
  \BibitemOpen
  \bibfield  {author} {\bibinfo {author} {\bibfnamefont {Andreas~P.}\
  \bibnamefont {Schnyder}}, \bibinfo {author} {\bibfnamefont {Shinsei}\
  \bibnamefont {Ryu}}, \bibinfo {author} {\bibfnamefont {Akira}\ \bibnamefont
  {Furusaki}}, \ and\ \bibinfo {author} {\bibfnamefont {Andreas W.~W.}\
  \bibnamefont {Ludwig}},\ }\bibfield  {title} {\enquote {\bibinfo {title}
  {Classification of topological insulators and superconductors in three
  spatial dimensions},}\ }\href {\doibase 10.1103/PhysRevB.78.195125}
  {\bibfield  {journal} {\bibinfo  {journal} {Phys. Rev. B}\ }\textbf {\bibinfo
  {volume} {78}},\ \bibinfo {pages} {195125} (\bibinfo {year}
  {2008})}\BibitemShut {NoStop}%
\bibitem [{\citenamefont {{Ryu}}\ \emph {et~al.}(2010)\citenamefont {{Ryu}},
  \citenamefont {{Schnyder}}, \citenamefont {{Furusaki}},\ and\ \citenamefont
  {{Ludwig}}}]{Ludwig2010}%
  \BibitemOpen
  \bibfield  {author} {\bibinfo {author} {\bibfnamefont {Shinsei}\ \bibnamefont
  {{Ryu}}}, \bibinfo {author} {\bibfnamefont {Andreas~P.}\ \bibnamefont
  {{Schnyder}}}, \bibinfo {author} {\bibfnamefont {Akira}\ \bibnamefont
  {{Furusaki}}}, \ and\ \bibinfo {author} {\bibfnamefont {Andreas W.~W.}\
  \bibnamefont {{Ludwig}}},\ }\bibfield  {title} {\enquote {\bibinfo {title}
  {{Topological insulators and superconductors: tenfold way and dimensional
  hierarchy}},}\ }\href {\doibase 10.1088/1367-2630/12/6/065010} {\bibfield
  {journal} {\bibinfo  {journal} {New Journal of Physics}\ }\textbf {\bibinfo
  {volume} {12}},\ \bibinfo {eid} {065010} (\bibinfo {year} {2010})},\ \Eprint
  {http://arxiv.org/abs/0912.2157} {arXiv:0912.2157 [cond-mat.mes-hall]}
  \BibitemShut {NoStop}%
\bibitem [{\citenamefont {Wu}\ \emph {et~al.}(2006)\citenamefont {Wu},
  \citenamefont {Bernevig},\ and\ \citenamefont {Zhang}}]{Zhang2006}%
  \BibitemOpen
  \bibfield  {author} {\bibinfo {author} {\bibfnamefont {Congjun}\ \bibnamefont
  {Wu}}, \bibinfo {author} {\bibfnamefont {B.~Andrei}\ \bibnamefont
  {Bernevig}}, \ and\ \bibinfo {author} {\bibfnamefont {Shou-Cheng}\
  \bibnamefont {Zhang}},\ }\bibfield  {title} {\enquote {\bibinfo {title}
  {Helical liquid and the edge of quantum spin hall systems},}\ }\href
  {\doibase 10.1103/PhysRevLett.96.106401} {\bibfield  {journal} {\bibinfo
  {journal} {Phys. Rev. Lett.}\ }\textbf {\bibinfo {volume} {96}},\ \bibinfo
  {pages} {106401} (\bibinfo {year} {2006})}\BibitemShut {NoStop}%
\bibitem [{\citenamefont {Xu}\ and\ \citenamefont {Moore}(2006)}]{Moore2006}%
  \BibitemOpen
  \bibfield  {author} {\bibinfo {author} {\bibfnamefont {Cenke}\ \bibnamefont
  {Xu}}\ and\ \bibinfo {author} {\bibfnamefont {J.~E.}\ \bibnamefont {Moore}},\
  }\bibfield  {title} {\enquote {\bibinfo {title} {Stability of the quantum
  spin hall effect: Effects of interactions, disorder, and ${\mathbb{z}}_{2}$
  topology},}\ }\href {\doibase 10.1103/PhysRevB.73.045322} {\bibfield
  {journal} {\bibinfo  {journal} {Phys. Rev. B}\ }\textbf {\bibinfo {volume}
  {73}},\ \bibinfo {pages} {045322} (\bibinfo {year} {2006})}\BibitemShut
  {NoStop}%
\bibitem [{\citenamefont {Nandkishore}\ and\ \citenamefont
  {Huse}(2015)}]{Huse2015}%
  \BibitemOpen
  \bibfield  {author} {\bibinfo {author} {\bibfnamefont {Rahul}\ \bibnamefont
  {Nandkishore}}\ and\ \bibinfo {author} {\bibfnamefont {David~A.}\
  \bibnamefont {Huse}},\ }\bibfield  {title} {\enquote {\bibinfo {title}
  {Many-body localization and thermalization in quantum statistical
  mechanics},}\ }\href {\doibase 10.1146/annurev-conmatphys-031214-014726}
  {\bibfield  {journal} {\bibinfo  {journal} {Annual Review of Condensed Matter
  Physics}\ }\textbf {\bibinfo {volume} {6}},\ \bibinfo {pages} {15--38}
  (\bibinfo {year} {2015})},\ \Eprint
  {http://arxiv.org/abs/https://doi.org/10.1146/annurev-conmatphys-031214-014726}
  {https://doi.org/10.1146/annurev-conmatphys-031214-014726} \BibitemShut
  {NoStop}%
\bibitem [{\citenamefont {Abanin}\ \emph {et~al.}(2019)\citenamefont {Abanin},
  \citenamefont {Altman}, \citenamefont {Bloch},\ and\ \citenamefont
  {Serbyn}}]{Serbyn2019}%
  \BibitemOpen
  \bibfield  {author} {\bibinfo {author} {\bibfnamefont {Dmitry~A.}\
  \bibnamefont {Abanin}}, \bibinfo {author} {\bibfnamefont {Ehud}\ \bibnamefont
  {Altman}}, \bibinfo {author} {\bibfnamefont {Immanuel}\ \bibnamefont
  {Bloch}}, \ and\ \bibinfo {author} {\bibfnamefont {Maksym}\ \bibnamefont
  {Serbyn}},\ }\bibfield  {title} {\enquote {\bibinfo {title} {Colloquium:
  Many-body localization, thermalization, and entanglement},}\ }\href {\doibase
  10.1103/RevModPhys.91.021001} {\bibfield  {journal} {\bibinfo  {journal}
  {Rev. Mod. Phys.}\ }\textbf {\bibinfo {volume} {91}},\ \bibinfo {pages}
  {021001} (\bibinfo {year} {2019})}\BibitemShut {NoStop}%
\bibitem [{\citenamefont {Parameswaran}\ and\ \citenamefont
  {Gopalakrishnan}(2017)}]{Gopalakrishnan2017}%
  \BibitemOpen
  \bibfield  {author} {\bibinfo {author} {\bibfnamefont {S.~A.}\ \bibnamefont
  {Parameswaran}}\ and\ \bibinfo {author} {\bibfnamefont {S.}~\bibnamefont
  {Gopalakrishnan}},\ }\bibfield  {title} {\enquote {\bibinfo {title}
  {Non-fermi glasses: Localized descendants of fractionalized metals},}\ }\href
  {\doibase 10.1103/PhysRevLett.119.146601} {\bibfield  {journal} {\bibinfo
  {journal} {Phys. Rev. Lett.}\ }\textbf {\bibinfo {volume} {119}},\ \bibinfo
  {pages} {146601} (\bibinfo {year} {2017})}\BibitemShut {NoStop}%
\bibitem [{\citenamefont {Fisher}\ \emph {et~al.}(1989)\citenamefont {Fisher},
  \citenamefont {Weichman}, \citenamefont {Grinstein},\ and\ \citenamefont
  {Fisher}}]{Fisher1989}%
  \BibitemOpen
  \bibfield  {author} {\bibinfo {author} {\bibfnamefont {Matthew P.~A.}\
  \bibnamefont {Fisher}}, \bibinfo {author} {\bibfnamefont {Peter~B.}\
  \bibnamefont {Weichman}}, \bibinfo {author} {\bibfnamefont {G.}~\bibnamefont
  {Grinstein}}, \ and\ \bibinfo {author} {\bibfnamefont {Daniel~S.}\
  \bibnamefont {Fisher}},\ }\bibfield  {title} {\enquote {\bibinfo {title}
  {Boson localization and the superfluid-insulator transition},}\ }\href
  {\doibase 10.1103/PhysRevB.40.546} {\bibfield  {journal} {\bibinfo  {journal}
  {Phys. Rev. B}\ }\textbf {\bibinfo {volume} {40}},\ \bibinfo {pages}
  {546--570} (\bibinfo {year} {1989})}\BibitemShut {NoStop}%
\bibitem [{\citenamefont {Serbyn}\ \emph {et~al.}(2013)\citenamefont {Serbyn},
  \citenamefont {Papi\ifmmode~\acute{c}\else \'{c}\fi{}},\ and\ \citenamefont
  {Abanin}}]{Abanin2013}%
  \BibitemOpen
  \bibfield  {author} {\bibinfo {author} {\bibfnamefont {Maksym}\ \bibnamefont
  {Serbyn}}, \bibinfo {author} {\bibfnamefont {Z.}~\bibnamefont
  {Papi\ifmmode~\acute{c}\else \'{c}\fi{}}}, \ and\ \bibinfo {author}
  {\bibfnamefont {Dmitry~A.}\ \bibnamefont {Abanin}},\ }\bibfield  {title}
  {\enquote {\bibinfo {title} {Local conservation laws and the structure of the
  many-body localized states},}\ }\href {\doibase
  10.1103/PhysRevLett.111.127201} {\bibfield  {journal} {\bibinfo  {journal}
  {Phys. Rev. Lett.}\ }\textbf {\bibinfo {volume} {111}},\ \bibinfo {pages}
  {127201} (\bibinfo {year} {2013})}\BibitemShut {NoStop}%
\bibitem [{\citenamefont {Huse}\ \emph {et~al.}(2014)\citenamefont {Huse},
  \citenamefont {Nandkishore},\ and\ \citenamefont
  {Oganesyan}}]{Oganesyan2014}%
  \BibitemOpen
  \bibfield  {author} {\bibinfo {author} {\bibfnamefont {David~A.}\
  \bibnamefont {Huse}}, \bibinfo {author} {\bibfnamefont {Rahul}\ \bibnamefont
  {Nandkishore}}, \ and\ \bibinfo {author} {\bibfnamefont {Vadim}\ \bibnamefont
  {Oganesyan}},\ }\bibfield  {title} {\enquote {\bibinfo {title} {Phenomenology
  of fully many-body-localized systems},}\ }\href {\doibase
  10.1103/PhysRevB.90.174202} {\bibfield  {journal} {\bibinfo  {journal} {Phys.
  Rev. B}\ }\textbf {\bibinfo {volume} {90}},\ \bibinfo {pages} {174202}
  (\bibinfo {year} {2014})}\BibitemShut {NoStop}%
\bibitem [{\citenamefont {{Imbrie}}(2016)}]{Imbrie2016}%
  \BibitemOpen
  \bibfield  {author} {\bibinfo {author} {\bibfnamefont {John~Z.}\ \bibnamefont
  {{Imbrie}}},\ }\bibfield  {title} {\enquote {\bibinfo {title} {{On Many-Body
  Localization for Quantum Spin Chains}},}\ }\href {\doibase
  10.1007/s10955-016-1508-x} {\bibfield  {journal} {\bibinfo  {journal}
  {Journal of Statistical Physics}\ }\textbf {\bibinfo {volume} {163}},\
  \bibinfo {pages} {998--1048} (\bibinfo {year} {2016})},\ \Eprint
  {http://arxiv.org/abs/1403.7837} {arXiv:1403.7837 [math-ph]} \BibitemShut
  {NoStop}%
\bibitem [{\citenamefont {Altshuler}\ \emph {et~al.}(2013)\citenamefont
  {Altshuler}, \citenamefont {Aleiner},\ and\ \citenamefont
  {Yudson}}]{Yudson2013}%
  \BibitemOpen
  \bibfield  {author} {\bibinfo {author} {\bibfnamefont {B.~L.}\ \bibnamefont
  {Altshuler}}, \bibinfo {author} {\bibfnamefont {I.~L.}\ \bibnamefont
  {Aleiner}}, \ and\ \bibinfo {author} {\bibfnamefont {V.~I.}\ \bibnamefont
  {Yudson}},\ }\bibfield  {title} {\enquote {\bibinfo {title} {Localization at
  the edge of a 2d topological insulator by kondo impurities with random
  anisotropies},}\ }\href {\doibase 10.1103/PhysRevLett.111.086401} {\bibfield
  {journal} {\bibinfo  {journal} {Phys. Rev. Lett.}\ }\textbf {\bibinfo
  {volume} {111}},\ \bibinfo {pages} {086401} (\bibinfo {year}
  {2013})}\BibitemShut {NoStop}%
\bibitem [{\citenamefont {Hsu}\ \emph {et~al.}(2017)\citenamefont {Hsu},
  \citenamefont {Stano}, \citenamefont {Klinovaja},\ and\ \citenamefont
  {Loss}}]{Loss2017}%
  \BibitemOpen
  \bibfield  {author} {\bibinfo {author} {\bibfnamefont {Chen-Hsuan}\
  \bibnamefont {Hsu}}, \bibinfo {author} {\bibfnamefont {Peter}\ \bibnamefont
  {Stano}}, \bibinfo {author} {\bibfnamefont {Jelena}\ \bibnamefont
  {Klinovaja}}, \ and\ \bibinfo {author} {\bibfnamefont {Daniel}\ \bibnamefont
  {Loss}},\ }\bibfield  {title} {\enquote {\bibinfo {title}
  {Nuclear-spin-induced localization of edge states in two-dimensional
  topological insulators},}\ }\href {\doibase 10.1103/PhysRevB.96.081405}
  {\bibfield  {journal} {\bibinfo  {journal} {Phys. Rev. B}\ }\textbf {\bibinfo
  {volume} {96}},\ \bibinfo {pages} {081405} (\bibinfo {year}
  {2017})}\BibitemShut {NoStop}%
\bibitem [{\citenamefont {Foster}\ and\ \citenamefont
  {Yuzbashyan}(2012)}]{Yuzbashyan2012}%
  \BibitemOpen
  \bibfield  {author} {\bibinfo {author} {\bibfnamefont {Matthew~S.}\
  \bibnamefont {Foster}}\ and\ \bibinfo {author} {\bibfnamefont {Emil~A.}\
  \bibnamefont {Yuzbashyan}},\ }\bibfield  {title} {\enquote {\bibinfo {title}
  {Interaction-mediated surface-state instability in disordered
  three-dimensional topological superconductors with spin su(2) symmetry},}\
  }\href {\doibase 10.1103/PhysRevLett.109.246801} {\bibfield  {journal}
  {\bibinfo  {journal} {Phys. Rev. Lett.}\ }\textbf {\bibinfo {volume} {109}},\
  \bibinfo {pages} {246801} (\bibinfo {year} {2012})}\BibitemShut {NoStop}%
\bibitem [{\citenamefont {Foster}\ \emph {et~al.}(2014)\citenamefont {Foster},
  \citenamefont {Xie},\ and\ \citenamefont {Chou}}]{Chou2014}%
  \BibitemOpen
  \bibfield  {author} {\bibinfo {author} {\bibfnamefont {Matthew~S.}\
  \bibnamefont {Foster}}, \bibinfo {author} {\bibfnamefont {Hong-Yi}\
  \bibnamefont {Xie}}, \ and\ \bibinfo {author} {\bibfnamefont {Yang-Zhi}\
  \bibnamefont {Chou}},\ }\bibfield  {title} {\enquote {\bibinfo {title}
  {Topological protection, disorder, and interactions: Survival at the surface
  of three-dimensional topological superconductors},}\ }\href {\doibase
  10.1103/PhysRevB.89.155140} {\bibfield  {journal} {\bibinfo  {journal} {Phys.
  Rev. B}\ }\textbf {\bibinfo {volume} {89}},\ \bibinfo {pages} {155140}
  (\bibinfo {year} {2014})}\BibitemShut {NoStop}%
\bibitem [{\citenamefont {Chou}\ \emph {et~al.}(2019)\citenamefont {Chou},
  \citenamefont {Nandkishore},\ and\ \citenamefont
  {Radzihovsky}}]{Radzihovsky2019}%
  \BibitemOpen
  \bibfield  {author} {\bibinfo {author} {\bibfnamefont {Yang-Zhi}\
  \bibnamefont {Chou}}, \bibinfo {author} {\bibfnamefont {Rahul~M.}\
  \bibnamefont {Nandkishore}}, \ and\ \bibinfo {author} {\bibfnamefont {Leo}\
  \bibnamefont {Radzihovsky}},\ }\bibfield  {title} {\enquote {\bibinfo {title}
  {Localized surfaces of three-dimensional topological insulators},}\ }\href
  {\doibase 10.1103/PhysRevB.99.165108} {\bibfield  {journal} {\bibinfo
  {journal} {Phys. Rev. B}\ }\textbf {\bibinfo {volume} {99}},\ \bibinfo
  {pages} {165108} (\bibinfo {year} {2019})}\BibitemShut {NoStop}%
\bibitem [{\citenamefont {Ran}\ \emph {et~al.}(2008)\citenamefont {Ran},
  \citenamefont {Vishwanath},\ and\ \citenamefont {Lee}}]{Lee2008}%
  \BibitemOpen
  \bibfield  {author} {\bibinfo {author} {\bibfnamefont {Ying}\ \bibnamefont
  {Ran}}, \bibinfo {author} {\bibfnamefont {Ashvin}\ \bibnamefont
  {Vishwanath}}, \ and\ \bibinfo {author} {\bibfnamefont {Dung-Hai}\
  \bibnamefont {Lee}},\ }\bibfield  {title} {\enquote {\bibinfo {title}
  {Spin-charge separated solitons in a topological band insulator},}\ }\href
  {\doibase 10.1103/PhysRevLett.101.086801} {\bibfield  {journal} {\bibinfo
  {journal} {Phys. Rev. Lett.}\ }\textbf {\bibinfo {volume} {101}},\ \bibinfo
  {pages} {086801} (\bibinfo {year} {2008})}\BibitemShut {NoStop}%
\bibitem [{\citenamefont {Qi}\ and\ \citenamefont {Zhang}(2008)}]{Zhang2008}%
  \BibitemOpen
  \bibfield  {author} {\bibinfo {author} {\bibfnamefont {Xiao-Liang}\
  \bibnamefont {Qi}}\ and\ \bibinfo {author} {\bibfnamefont {Shou-Cheng}\
  \bibnamefont {Zhang}},\ }\bibfield  {title} {\enquote {\bibinfo {title}
  {Spin-charge separation in the quantum spin hall state},}\ }\href {\doibase
  10.1103/PhysRevLett.101.086802} {\bibfield  {journal} {\bibinfo  {journal}
  {Phys. Rev. Lett.}\ }\textbf {\bibinfo {volume} {101}},\ \bibinfo {pages}
  {086802} (\bibinfo {year} {2008})}\BibitemShut {NoStop}%
\bibitem [{\citenamefont {Qi}\ \emph {et~al.}(2008)\citenamefont {Qi},
  \citenamefont {Hughes},\ and\ \citenamefont {Zhang}}]{Zhang2008a}%
  \BibitemOpen
  \bibfield  {author} {\bibinfo {author} {\bibfnamefont {Xiao-Liang}\
  \bibnamefont {Qi}}, \bibinfo {author} {\bibfnamefont {Taylor~L.}\
  \bibnamefont {Hughes}}, \ and\ \bibinfo {author} {\bibfnamefont {Shou-Cheng}\
  \bibnamefont {Zhang}},\ }\bibfield  {title} {\enquote {\bibinfo {title}
  {Fractional charge and quantized current in the quantum spin hall state},}\
  }\href {https://doi.org/10.1038/nphys913} {\bibfield  {journal} {\bibinfo
  {journal} {Nature Physics}\ }\textbf {\bibinfo {volume} {4}},\ \bibinfo
  {pages} {273} (\bibinfo {year} {2008})}\BibitemShut {NoStop}%
\bibitem [{\citenamefont {Chen}\ and\ \citenamefont
  {Vishwanath}(2015)}]{Vishwanath2015}%
  \BibitemOpen
  \bibfield  {author} {\bibinfo {author} {\bibfnamefont {Xie}\ \bibnamefont
  {Chen}}\ and\ \bibinfo {author} {\bibfnamefont {Ashvin}\ \bibnamefont
  {Vishwanath}},\ }\bibfield  {title} {\enquote {\bibinfo {title} {Towards
  gauging time-reversal symmetry: A tensor network approach},}\ }\href
  {\doibase 10.1103/PhysRevX.5.041034} {\bibfield  {journal} {\bibinfo
  {journal} {Phys. Rev. X}\ }\textbf {\bibinfo {volume} {5}},\ \bibinfo {pages}
  {041034} (\bibinfo {year} {2015})}\BibitemShut {NoStop}%
\bibitem [{\citenamefont {Potter}\ and\ \citenamefont
  {Vasseur}(2016)}]{Vasseur2016}%
  \BibitemOpen
  \bibfield  {author} {\bibinfo {author} {\bibfnamefont {Andrew~C.}\
  \bibnamefont {Potter}}\ and\ \bibinfo {author} {\bibfnamefont {Romain}\
  \bibnamefont {Vasseur}},\ }\bibfield  {title} {\enquote {\bibinfo {title}
  {Symmetry constraints on many-body localization},}\ }\href {\doibase
  10.1103/PhysRevB.94.224206} {\bibfield  {journal} {\bibinfo  {journal} {Phys.
  Rev. B}\ }\textbf {\bibinfo {volume} {94}},\ \bibinfo {pages} {224206}
  (\bibinfo {year} {2016})}\BibitemShut {NoStop}%
\bibitem [{\citenamefont {Huse}\ \emph {et~al.}(2013)\citenamefont {Huse},
  \citenamefont {Nandkishore}, \citenamefont {Oganesyan}, \citenamefont {Pal},\
  and\ \citenamefont {Sondhi}}]{Sondhi2013}%
  \BibitemOpen
  \bibfield  {author} {\bibinfo {author} {\bibfnamefont {David~A.}\
  \bibnamefont {Huse}}, \bibinfo {author} {\bibfnamefont {Rahul}\ \bibnamefont
  {Nandkishore}}, \bibinfo {author} {\bibfnamefont {Vadim}\ \bibnamefont
  {Oganesyan}}, \bibinfo {author} {\bibfnamefont {Arijeet}\ \bibnamefont
  {Pal}}, \ and\ \bibinfo {author} {\bibfnamefont {S.~L.}\ \bibnamefont
  {Sondhi}},\ }\bibfield  {title} {\enquote {\bibinfo {title}
  {Localization-protected quantum order},}\ }\href {\doibase
  10.1103/PhysRevB.88.014206} {\bibfield  {journal} {\bibinfo  {journal} {Phys.
  Rev. B}\ }\textbf {\bibinfo {volume} {88}},\ \bibinfo {pages} {014206}
  (\bibinfo {year} {2013})}\BibitemShut {NoStop}%
\bibitem [{\citenamefont {De~Roeck}\ \emph {et~al.}(2016)\citenamefont
  {De~Roeck}, \citenamefont {Huveneers}, \citenamefont {M\"uller},\ and\
  \citenamefont {Schiulaz}}]{Schiulaz2016}%
  \BibitemOpen
  \bibfield  {author} {\bibinfo {author} {\bibfnamefont {Wojciech}\
  \bibnamefont {De~Roeck}}, \bibinfo {author} {\bibfnamefont {Francois}\
  \bibnamefont {Huveneers}}, \bibinfo {author} {\bibfnamefont {Markus}\
  \bibnamefont {M\"uller}}, \ and\ \bibinfo {author} {\bibfnamefont {Mauro}\
  \bibnamefont {Schiulaz}},\ }\bibfield  {title} {\enquote {\bibinfo {title}
  {Absence of many-body mobility edges},}\ }\href {\doibase
  10.1103/PhysRevB.93.014203} {\bibfield  {journal} {\bibinfo  {journal} {Phys.
  Rev. B}\ }\textbf {\bibinfo {volume} {93}},\ \bibinfo {pages} {014203}
  (\bibinfo {year} {2016})}\BibitemShut {NoStop}%
\bibitem [{\citenamefont {Imry}\ and\ \citenamefont {Ma}(1975)}]{Ma1975}%
  \BibitemOpen
  \bibfield  {author} {\bibinfo {author} {\bibfnamefont {Yoseph}\ \bibnamefont
  {Imry}}\ and\ \bibinfo {author} {\bibfnamefont {Shang-keng}\ \bibnamefont
  {Ma}},\ }\bibfield  {title} {\enquote {\bibinfo {title} {Random-field
  instability of the ordered state of continuous symmetry},}\ }\href {\doibase
  10.1103/PhysRevLett.35.1399} {\bibfield  {journal} {\bibinfo  {journal}
  {Phys. Rev. Lett.}\ }\textbf {\bibinfo {volume} {35}},\ \bibinfo {pages}
  {1399--1401} (\bibinfo {year} {1975})}\BibitemShut {NoStop}%
\bibitem [{\citenamefont {Binder}(1983)}]{Binder1983}%
  \BibitemOpen
  \bibfield  {author} {\bibinfo {author} {\bibfnamefont {K.}~\bibnamefont
  {Binder}},\ }\bibfield  {title} {\enquote {\bibinfo {title} {Random-field
  induced interface widths in ising systems},}\ }\href {\doibase
  10.1007/BF01470045} {\bibfield  {journal} {\bibinfo  {journal} {Zeitschrift
  f{\"u}r Physik B Condensed Matter}\ }\textbf {\bibinfo {volume} {50}},\
  \bibinfo {pages} {343--352} (\bibinfo {year} {1983})}\BibitemShut {NoStop}%
\bibitem [{\citenamefont {Langer}\ and\ \citenamefont
  {Fisher}(1967)}]{Fisher1967}%
  \BibitemOpen
  \bibfield  {author} {\bibinfo {author} {\bibfnamefont {J.~S.}\ \bibnamefont
  {Langer}}\ and\ \bibinfo {author} {\bibfnamefont {Michael~E.}\ \bibnamefont
  {Fisher}},\ }\bibfield  {title} {\enquote {\bibinfo {title} {Intrinsic
  critical velocity of a superfluid},}\ }\href {\doibase
  10.1103/PhysRevLett.19.560} {\bibfield  {journal} {\bibinfo  {journal} {Phys.
  Rev. Lett.}\ }\textbf {\bibinfo {volume} {19}},\ \bibinfo {pages} {560--563}
  (\bibinfo {year} {1967})}\BibitemShut {NoStop}%
\bibitem [{\citenamefont {Langer}\ and\ \citenamefont
  {Ambegaokar}(1967)}]{Ambegaokar1967}%
  \BibitemOpen
  \bibfield  {author} {\bibinfo {author} {\bibfnamefont {J.~S.}\ \bibnamefont
  {Langer}}\ and\ \bibinfo {author} {\bibfnamefont {Vinay}\ \bibnamefont
  {Ambegaokar}},\ }\bibfield  {title} {\enquote {\bibinfo {title} {Intrinsic
  resistive transition in narrow superconducting channels},}\ }\href {\doibase
  10.1103/PhysRev.164.498} {\bibfield  {journal} {\bibinfo  {journal} {Phys.
  Rev.}\ }\textbf {\bibinfo {volume} {164}},\ \bibinfo {pages} {498--510}
  (\bibinfo {year} {1967})}\BibitemShut {NoStop}%
\bibitem [{\citenamefont {McCumber}\ and\ \citenamefont
  {Halperin}(1970)}]{Halperin1970}%
  \BibitemOpen
  \bibfield  {author} {\bibinfo {author} {\bibfnamefont {D.~E.}\ \bibnamefont
  {McCumber}}\ and\ \bibinfo {author} {\bibfnamefont {B.~I.}\ \bibnamefont
  {Halperin}},\ }\bibfield  {title} {\enquote {\bibinfo {title} {Time scale of
  intrinsic resistive fluctuations in thin superconducting wires},}\ }\href
  {\doibase 10.1103/PhysRevB.1.1054} {\bibfield  {journal} {\bibinfo  {journal}
  {Phys. Rev. B}\ }\textbf {\bibinfo {volume} {1}},\ \bibinfo {pages}
  {1054--1070} (\bibinfo {year} {1970})}\BibitemShut {NoStop}%
\bibitem [{\citenamefont {Fisher}\ and\ \citenamefont {Huse}(1986)}]{Huse1986}%
  \BibitemOpen
  \bibfield  {author} {\bibinfo {author} {\bibfnamefont {Daniel~S.}\
  \bibnamefont {Fisher}}\ and\ \bibinfo {author} {\bibfnamefont {David~A.}\
  \bibnamefont {Huse}},\ }\bibfield  {title} {\enquote {\bibinfo {title}
  {Ordered phase of short-range ising spin-glasses},}\ }\href {\doibase
  10.1103/PhysRevLett.56.1601} {\bibfield  {journal} {\bibinfo  {journal}
  {Phys. Rev. Lett.}\ }\textbf {\bibinfo {volume} {56}},\ \bibinfo {pages}
  {1601--1604} (\bibinfo {year} {1986})}\BibitemShut {NoStop}%
\bibitem [{\citenamefont {Fisher}\ \emph {et~al.}(1991)\citenamefont {Fisher},
  \citenamefont {Fisher},\ and\ \citenamefont {Huse}}]{Huse1991}%
  \BibitemOpen
  \bibfield  {author} {\bibinfo {author} {\bibfnamefont {Daniel~S.}\
  \bibnamefont {Fisher}}, \bibinfo {author} {\bibfnamefont {Matthew P.~A.}\
  \bibnamefont {Fisher}}, \ and\ \bibinfo {author} {\bibfnamefont {David~A.}\
  \bibnamefont {Huse}},\ }\bibfield  {title} {\enquote {\bibinfo {title}
  {Thermal fluctuations, quenched disorder, phase transitions, and transport in
  type-ii superconductors},}\ }\href {\doibase 10.1103/PhysRevB.43.130}
  {\bibfield  {journal} {\bibinfo  {journal} {Phys. Rev. B}\ }\textbf {\bibinfo
  {volume} {43}},\ \bibinfo {pages} {130--159} (\bibinfo {year}
  {1991})}\BibitemShut {NoStop}%
\bibitem [{\citenamefont {{Balram}}\ \emph {et~al.}(2019)\citenamefont
  {{Balram}}, \citenamefont {{Flensberg}}, \citenamefont {{Paaske}},\ and\
  \citenamefont {{Rudner}}}]{Rudner2019}%
  \BibitemOpen
  \bibfield  {author} {\bibinfo {author} {\bibfnamefont {Ajit~C.}\ \bibnamefont
  {{Balram}}}, \bibinfo {author} {\bibfnamefont {Karsten}\ \bibnamefont
  {{Flensberg}}}, \bibinfo {author} {\bibfnamefont {Jens}\ \bibnamefont
  {{Paaske}}}, \ and\ \bibinfo {author} {\bibfnamefont {Mark~S.}\ \bibnamefont
  {{Rudner}}},\ }\bibfield  {title} {\enquote {\bibinfo {title}
  {{Current-induced gap opening in interacting topological insulator
  surfaces}},}\ }\href@noop {} {\bibfield  {journal} {\bibinfo  {journal}
  {arXiv e-prints}\ ,\ \bibinfo {eid} {arXiv:1901.08067}} (\bibinfo {year}
  {2019})},\ \Eprint {http://arxiv.org/abs/1901.08067} {arXiv:1901.08067
  [cond-mat.mes-hall]} \BibitemShut {NoStop}%
\bibitem [{\citenamefont {{Goss Levi}}(1998)}]{GossLevi1998}%
  \BibitemOpen
  \bibfield  {author} {\bibinfo {author} {\bibfnamefont {B.}~\bibnamefont
  {{Goss Levi}}},\ }\bibfield  {title} {\enquote {\bibinfo {title} {{Scanning
  Microscopes Probe Local Details of the Quantum Hall State}},}\ }\href@noop {}
  {\bibfield  {journal} {\bibinfo  {journal} {Physics Today}\ }\textbf
  {\bibinfo {volume} {51}},\ \bibinfo {pages} {17--19} (\bibinfo {year}
  {1998})}\BibitemShut {NoStop}%
\bibitem [{\citenamefont {Swingle}\ and\ \citenamefont {Yao}(2017)}]{Yao2017}%
  \BibitemOpen
  \bibfield  {author} {\bibinfo {author} {\bibfnamefont {Brian}\ \bibnamefont
  {Swingle}}\ and\ \bibinfo {author} {\bibfnamefont {Norman~Y}\ \bibnamefont
  {Yao}},\ }\bibfield  {title} {\enquote {\bibinfo {title} {Seeing scrambled
  spins},}\ }\href@noop {} {\bibfield  {journal} {\bibinfo  {journal}
  {Physics}\ }\textbf {\bibinfo {volume} {10}},\ \bibinfo {pages} {82}
  (\bibinfo {year} {2017})}\BibitemShut {NoStop}%
\bibitem [{\citenamefont {Cheng}\ \emph {et~al.}(2016)\citenamefont {Cheng},
  \citenamefont {Zaletel}, \citenamefont {Barkeshli}, \citenamefont
  {Vishwanath},\ and\ \citenamefont {Bonderson}}]{Bonderson2016}%
  \BibitemOpen
  \bibfield  {author} {\bibinfo {author} {\bibfnamefont {Meng}\ \bibnamefont
  {Cheng}}, \bibinfo {author} {\bibfnamefont {Michael}\ \bibnamefont
  {Zaletel}}, \bibinfo {author} {\bibfnamefont {Maissam}\ \bibnamefont
  {Barkeshli}}, \bibinfo {author} {\bibfnamefont {Ashvin}\ \bibnamefont
  {Vishwanath}}, \ and\ \bibinfo {author} {\bibfnamefont {Parsa}\ \bibnamefont
  {Bonderson}},\ }\bibfield  {title} {\enquote {\bibinfo {title} {Translational
  symmetry and microscopic constraints on symmetry-enriched topological phases:
  A view from the surface},}\ }\href {\doibase 10.1103/PhysRevX.6.041068}
  {\bibfield  {journal} {\bibinfo  {journal} {Phys. Rev. X}\ }\textbf {\bibinfo
  {volume} {6}},\ \bibinfo {pages} {041068} (\bibinfo {year}
  {2016})}\BibitemShut {NoStop}%
\bibitem [{\citenamefont {{Lieb}}\ \emph {et~al.}(1961)\citenamefont {{Lieb}},
  \citenamefont {{Schultz}},\ and\ \citenamefont {{Mattis}}}]{Mattis1961}%
  \BibitemOpen
  \bibfield  {author} {\bibinfo {author} {\bibfnamefont {E.}~\bibnamefont
  {{Lieb}}}, \bibinfo {author} {\bibfnamefont {T.}~\bibnamefont {{Schultz}}}, \
  and\ \bibinfo {author} {\bibfnamefont {D.}~\bibnamefont {{Mattis}}},\
  }\bibfield  {title} {\enquote {\bibinfo {title} {{Two soluble models of an
  antiferromagnetic chain}},}\ }\href {\doibase 10.1016/0003-4916(61)90115-4}
  {\bibfield  {journal} {\bibinfo  {journal} {Ann. Phys.}\ }\textbf {\bibinfo
  {volume} {16}},\ \bibinfo {pages} {407--466} (\bibinfo {year}
  {1961})}\BibitemShut {NoStop}%
\bibitem [{\citenamefont {Oshikawa}(2000)}]{Oshikawa2000}%
  \BibitemOpen
  \bibfield  {author} {\bibinfo {author} {\bibfnamefont {Masaki}\ \bibnamefont
  {Oshikawa}},\ }\bibfield  {title} {\enquote {\bibinfo {title}
  {Commensurability, excitation gap, and topology in quantum many-particle
  systems on a periodic lattice},}\ }\href {\doibase
  10.1103/PhysRevLett.84.1535} {\bibfield  {journal} {\bibinfo  {journal}
  {Phys. Rev. Lett.}\ }\textbf {\bibinfo {volume} {84}},\ \bibinfo {pages}
  {1535--1538} (\bibinfo {year} {2000})}\BibitemShut {NoStop}%
\bibitem [{\citenamefont {Kimchi}\ \emph {et~al.}(2018)\citenamefont {Kimchi},
  \citenamefont {Nahum},\ and\ \citenamefont {Senthil}}]{randommagnets}%
  \BibitemOpen
  \bibfield  {author} {\bibinfo {author} {\bibfnamefont {Itamar}\ \bibnamefont
  {Kimchi}}, \bibinfo {author} {\bibfnamefont {Adam}\ \bibnamefont {Nahum}}, \
  and\ \bibinfo {author} {\bibfnamefont {T.}~\bibnamefont {Senthil}},\
  }\bibfield  {title} {\enquote {\bibinfo {title} {Valence bonds in random
  quantum magnets: Theory and application to {YbMgGaO}$_{4}$},}\ }\href
  {\doibase 10.1103/PhysRevX.8.031028} {\bibfield  {journal} {\bibinfo
  {journal} {Phys. Rev. X}\ }\textbf {\bibinfo {volume} {8}},\ \bibinfo {pages}
  {031028} (\bibinfo {year} {2018})}\BibitemShut {NoStop}%
\bibitem [{\citenamefont {Fisher}(1994)}]{Fisher1994}%
  \BibitemOpen
  \bibfield  {author} {\bibinfo {author} {\bibfnamefont {Daniel~S.}\
  \bibnamefont {Fisher}},\ }\bibfield  {title} {\enquote {\bibinfo {title}
  {Random antiferromagnetic quantum spin chains},}\ }\href {\doibase
  10.1103/PhysRevB.50.3799} {\bibfield  {journal} {\bibinfo  {journal} {Phys.
  Rev. B}\ }\textbf {\bibinfo {volume} {50}},\ \bibinfo {pages} {3799--3821}
  (\bibinfo {year} {1994})}\BibitemShut {NoStop}%
\bibitem [{\citenamefont {{Metlitski}}(2019)}]{Metlitski2019}%
  \BibitemOpen
  \bibfield  {author} {\bibinfo {author} {\bibfnamefont {Max~A.}\ \bibnamefont
  {{Metlitski}}},\ }\bibfield  {title} {\enquote {\bibinfo {title} {{A 1d
  lattice model for the boundary of the quantum spin-Hall insulator}},}\
  }\href@noop {} {\bibfield  {journal} {\bibinfo  {journal} {arXiv e-prints}\
  ,\ \bibinfo {eid} {arXiv:1908.08958}} (\bibinfo {year} {2019})},\ \Eprint
  {http://arxiv.org/abs/1908.08958} {arXiv:1908.08958 [cond-mat.str-el]}
  \BibitemShut {NoStop}%
\bibitem [{\citenamefont {{Son}}\ and\ \citenamefont
  {{Alicea}}(2019)}]{Alicea2019}%
  \BibitemOpen
  \bibfield  {author} {\bibinfo {author} {\bibfnamefont {Jun~Ho}\ \bibnamefont
  {{Son}}}\ and\ \bibinfo {author} {\bibfnamefont {Jason}\ \bibnamefont
  {{Alicea}}},\ }\bibfield  {title} {\enquote {\bibinfo {title}
  {{Commuting-projector Hamiltonians for 2D topological insulators: edge
  physics and many-body invariants}},}\ }\href@noop {} {\bibfield  {journal}
  {\bibinfo  {journal} {arXiv e-prints}\ ,\ \bibinfo {eid} {arXiv:1906.11846}}
  (\bibinfo {year} {2019})},\ \Eprint {http://arxiv.org/abs/1906.11846}
  {arXiv:1906.11846 [cond-mat.str-el]} \BibitemShut {NoStop}%
\end{thebibliography}%

\end{document}